\documentclass[number,preprint,3p]{elsarticle}

\usepackage{multirow}
\usepackage{color}
\usepackage{graphicx}
\usepackage{subcaption}
\usepackage{algorithm}
\usepackage{bm}
\usepackage[colorlinks]{hyperref}
\usepackage{amssymb}
\usepackage{amsthm}
\usepackage{amsmath}
\usepackage{booktabs}
\usepackage{url}
\usepackage{scrextend}
\usepackage{epstopdf}
\usepackage{float}
\usepackage{tablefootnote}
\usepackage{mathtools}
\usepackage[perpage]{footmisc}

\makeatletter
\newenvironment{breakablealgorithm}
  {
   \begin{center}
     \refstepcounter{algorithm}
     \hrule height.8pt depth0pt \kern2pt
     \renewcommand{\caption}[2][\relax]{
       {\raggedright\textbf{\ALG@name~\thealgorithm} ##2\par}%
       \ifx\relax##1\relax 
         \addcontentsline{loa}{algorithm}{\protect\numberline{\thealgorithm}##2}%
       \else 
         \addcontentsline{loa}{algorithm}{\protect\numberline{\thealgorithm}##1}%
       \fi
       \kern2pt\hrule\kern2pt
     }
  }{
     \kern2pt\hrule\relax
   \end{center}
  }
\makeatother

\newcommand{\Prob}[1]{\mathbb{P}\left(#1\right)}
\newcommand{\E}[1]{\mathbb{E}\left[#1\right]}

\newcommand{\vect}[1]{\boldsymbol{#1}}

\biboptions{square}
\journal{arXiv}
\begin{document}

\begin{frontmatter}
\renewcommand{\thefootnote}{\fnsymbol{footnote}}



\title{Adaptive active subspace-based metamodeling for high-dimensional reliability analysis}

 \author[1]{Jungho Kim}
 \author[1]{Ziqi Wang\footnotemark[1]}
 \author[2]{Junho Song\footnotemark[2]}
 \address[1]{Department of Civil and Environmental Engineering, University of California, Berkeley, CA, USA}
 \address[2]{Department of Civil and Environmental Engineering, Seoul National University, Seoul, Republic of Korea}
 \footnotetext[1]{Corresponding author: Ziqi Wang; \href{mailto:ziqiwang@berkeley.edu}{ziqiwang@berkeley.edu}}
 \footnotetext[2]{Corresponding author: Junho Song; \href{mailto:junhosong@snu.ac.kr}{junhosong@snu.ac.kr}}

\begin{abstract}
To address the challenges of reliability analysis in high-dimensional probability spaces, this paper proposes a new metamodeling method that couples active subspace, heteroscedastic Gaussian process, and active learning. The active subspace is leveraged to identify low-dimensional salient features of a high-dimensional computational model. A surrogate computational model is built in the low-dimensional feature space by a heteroscedastic Gaussian process. Active learning adaptively guides the surrogate model training toward the critical region that significantly contributes to the failure probability. A critical trait of the proposed method is that the three main ingredients–active subspace, heteroscedastic Gaussian process, and active learning–are coupled to adaptively optimize the feature space mapping in conjunction with the surrogate modeling. This coupling empowers the proposed method to accurately solve nontrivial high-dimensional reliability problems via low-dimensional surrogate modeling. Finally, numerical examples of a high-dimensional nonlinear function and structural engineering applications are investigated to verify the performance of the proposed method.
\end{abstract}

\begin{keyword}
active learning \sep active subspace \sep dimensionality reduction \sep high-dimensional \sep metamodeling \sep structural reliability

\end{keyword}

\end{frontmatter}

\renewcommand{\thefootnote}{\fnsymbol{footnote}}

\section{Introduction}

\noindent The performance evaluation of an engineering system is significantly affected by ubiquitous uncertainties arising from the lack of data, modeling approximations, and inherent randomness in the system and its environment. Therefore, quantifying and understanding the impact of uncertainties is essential. Forward uncertainty quantification (UQ) and reliability analysis are formal computational frameworks to analyze the impact of uncertainties on engineering systems. The former aims to obtain generic statistical quantities of interest (QoI), while the latter focuses on the probability of failure. This study addresses one of the unresolved challenges in forward UQ and reliability analysis–the efficient estimation of rare-event probabilities in high-dimensional spaces.

Generally, rare-event probability estimation requires numerous evaluations of computational models to solve a multidimensional integral. This challenge is exacerbated by (1) models involving complex and high-fidelity physics-based simulations and (2) many input random variables. The first issue makes Monte Carlo simulation (MCS) undesirable, and one should devise “clever” algorithms that require a minimum number of model evaluations. The second issue, however, favors MCS, which is an attractive (sometimes the only) option for generic high-dimensional integration. Several advanced variance-reduction techniques have been proposed for rare-event simulation, such as Subset Simulation \cite{au2003subset}\cite{zuev2012bayesian}\cite{wang2019hamiltonian} and Importance Sampling \cite{wang2016cross}\cite{papaioannou2019improved}. However, these advanced methods typically require thousands of model evaluations, limiting their applications to problems with expensive computational models.

Metamodeling approaches to address both issues showed limitations because high-dimensional metamodeling generally suffers from the curse of dimensionality \cite{lataniotis2020extending}\cite{kim2021quantile}\cite{kontolati2022survey}. A natural solution is metamodeling with dimensionality reduction techniques following the two steps, i.e., (1) mapping the input variables to a low-dimensional feature space/manifold and (2) constructing a surrogate model in the reduced feature space. Several approaches introduced for this purpose use observations of the input only, i.e., unsupervised learning without coupling with the computational model. For example, Kalogeris and Papadopoulos (2020) introduced diffusion map-based metamodeling that identifies the parameterization of a low-dimensional manifold by eigenfunctions of a diffusion operator \cite{kalogeris2020diffusion}. Giovanis and Shields (2018) proposed a method that combines manifold learning with a metamodel by projecting subspace-structured features onto a Grassmann manifold \cite{giovanis2018uncertainty}. The framework was further developed by combining with multipoint nonlinear kernel-based dimensionality reduction, called Grassmann diffusion maps \cite{giovanis2020data}\cite{dos2022grassmannian}. The unsupervised dimensionality reduction has straightforward implementation and could be effective for some applications. However, the approach shows significant limitations when the high-dimensional input lacks regularity, or the low-dimensional representation exhibits a complex topology unsuitable for metamodeling \cite{lataniotis2020extending}\cite{li2020deep}\cite{jiang2021recursive}.

On the other hand, as a supervised dimensionality reduction technique, the active subspace method identifies a low-dimensional feature space for the high-dimensional computational model by computing the maximal variation directions in the tangent space of the computational model output \cite{constantine2014active}. The tight coupling with a computational model helps the active subspace method overcome the limitations of unsupervised dimensionality reduction, but at the cost of gradient computations. Active subspace has been increasingly applied to forward UQ and reliability analysis. For example, Jiang and Li (2017) and Zhou and Peng (2021) employed an active subspace-based dimensionality reduction scheme to facilitate applications of the probability density evolution method to high-dimensional seismic reliability analysis \cite{jiang2017high}\cite{zhou2021active}. Navaneeth and Chakraborty (2022) proposed a high-dimensional reliability analysis framework that combines sparse polynomial chaos expansion with the active subspace algorithm \cite{navaneeth2022surrogate}.

Building on existing progress, this paper develops a new adaptive metamodeling method: adaptive active subspace-based heteroscedastic Gaussian process (AaS-hGP). In AaS-hGP, the low-dimensional features of a high-dimensional computational model are adaptively identified and optimized by the active subspace algorithm in conjunction with an active learning scheme at the outer loop. For a given reduced feature space, a heteroscedastic Gaussian process (hGP) is constructed as a low-dimensional surrogate of the original high-dimensional computational model. The hGP is desirable in the proposed method because it can accommodate prediction errors arising from the low-dimensional surrogate modeling in the reduced feature space. The coupling of dimensionality reduction, surrogate modeling, and active learning enables the AaS-hGP framework to achieve high computational efficiency and accuracy, thus offering an attractive option for high-dimensional reliability analysis and rare-event simulation.

The paper first briefly overviews the general formulations of high-dimensional reliability problems and dimensionality reduction-based metamodeling in Section \ref{Background}. Section \ref{AaS-hGP} sequentially introduces the following details of the proposed AaS-hGP framework: (1) identification of low-dimensional features by active subspace, (2) response predictions in the reduced feature space via hGP surrogate modeling, (3) active learning scheme, and (4) the algorithm of AaS-hGP. The numerical examples in Section \ref{Examples} demonstrate the performance of the proposed AaS-hGP method for high-dimensional reliability problems. Section \ref{Issues} discusses various practical issues, limitations, and future research topics regarding the proposed method. Lastly, a summary and concluding remarks are provided in Section \ref{Conclusion}.

\section{Background}\label{Background}

\subsection{High-dimensional reliability analysis}

\noindent Reliability is defined as the probability that a system remains functional, considering uncertainties from various sources \cite{der2022structural}\cite{song2021structural}. Consider a system with a $D$-dimensional vector of basic random variables $\vect{X}=[X_1,X_2,...,X_D]$, representing the source of uncertainties, and a performance measure $Y$ propagated from $\vect{X}$, representing the system’s tendency to fail. The performance measure $Y$ is written as
\begin{equation}
Y=\mathcal{M}\left(\vect{X}\right)\, ,\label{ModelF}
\end{equation}
\noindent where $\mathcal{M}:\vect{X}\in\mathbb{R}^D\mapsto{Y}\in\mathbb{R}$ denotes a general mathematical model that typically involves nonlinear physics-based simulations.

In a reliability problem, the failure probability $P_f$ is described as the $D$-fold integral
\begin{equation}
P_f=\Prob{\mathcal{M}\left(\vect{X}\right)\geq y_f}=\int_{\mathcal{M}\left(\vect{x}\right)\geq y_f}f_{\vect{X}}(\vect{x})\,d\vect{x}=\int_{\mathbb{R}^D}\mathbb{I}\left(\mathcal{M}\left(\vect{x}\right)\geq y_f\right)f_{\vect{X}}(\vect{x})\,d\vect{x}\, ,\label{Pf}
\end{equation}
\noindent where $\mathbb{P}(\cdot)$ denotes probability, $f_{\vect{X}}(\vect{x})$ denotes the joint probability density functions (PDF) of random vector $\vect{X}$, $\vect{x}$ is the realization of $\vect{X}$, $y_f$ is a prescribed threshold to define failure, and $\mathbb{I}\left(\mathcal{M}\left(\vect{x}\right)\geq y_f\right)$  denotes a binary indicator function that gives “1” if $\mathcal{M}\left(\vect{x}\right)\geq y_f$ and “0” otherwise. Notably, in the reliability community, the failure event $\mathcal{M}\left(\vect{x}\right)\geq y_f$ is often represented by $G\left(\vect{x}\right)\leq0$, where $G(\cdot)$ is the limit-state function. We adopt the current notations to highlight the close link of reliability to UQ formulations with a generic computational model $\mathcal{M}$. Solving Eq.\eqref{Pf} typically requires a substantial number of simulations of the model $\mathcal{M}$, and in general, it becomes increasingly challenging \cite{alibrandi2014response}\cite{kim2023estimation} as the dimensionality of $\vect{X}$ grows.

\subsection{Dimensionality reduction and metamodeling}

\noindent The purpose of dimensionality reduction is to represent the high-dimensional vector $\vect{X}\in\mathbb{R}^D$ by a low-dimensional feature vector $\vect{\Psi}\in\mathbb{R}^{d_r}$. The dimensionality reduction can be represented as
\begin{equation}
\vect{\Psi}=\mathcal{H}\left(\vect{X};\vect{\theta}_{\mathcal{H}}\right),\,\,\,\,\, \hat{\vect{X}}=\mathcal{H}^{-1}\left(\vect{\Psi};\vect{\theta}_{\mathcal{H}}\right)\, ,\label{DR}
\end{equation}
\noindent where $\mathcal{H}:\vect{X}\in\mathbb{R}^D\mapsto\vect{\Psi}\in\mathbb{R}^{d_r}$ denotes the dimensionality reduction function, $\vect{\theta}_{\mathcal{H}}$ denotes parameters that characterize the dimensionality reduction, $D$ and $d_r$ respectively denote the dimensions of vectors $\vect{X}$ and $\vect{\Psi}$, and $\mathcal{H}^{-1}:\vect{\Psi}\in\mathbb{R}^{d_r}\mapsto\hat{\vect{X}}\in\mathbb{R}^D$ represents the reconstruction function. Note that the reconstruction function $\mathcal{H}^{-1}$ is often approximate since the dimensionality reduction mapping is generally not bijective. Several approaches to construct $\mathcal{H}(\cdot)$ have been used for reliability analysis and optimization (see Ref. \cite{van2009dimensionality} for a comprehensive review); they can be categorized into linear or nonlinear dimensionality reduction methods. Examples of linear methods include principal component analysis, multidimensional scaling, Fisher’s linear discriminant analysis, and locality-preserving projections; examples of nonlinear methods include kernel principal component analysis, Laplacian eigenmaps, diffusion maps, Isomap, locally-linear embedding, and autoencoders. The optimal parameters of dimensionality reduction are often obtained by minimizing a distance measure between the original $\vect{X}$ and its reconstruction $\hat{\vect{X}}=\mathcal{H}^{-1}\left(\mathcal{H}\left(\vect{X}\right)\right)$.

The motivation of metamodeling \cite{blatman2011adaptive}\cite{zhang2019accelerating}\cite{kim2020probability} is straightforward: if an expensive computational model can be effectively replaced by a cheap surrogate, reliability and UQ analysis, along with other outer-loop simulations, will be sped up significantly. The metamodel of $Y=\mathcal{M}\left(\vect{X}\right)$ constructed in the reduced feature space can be written as
\begin{equation}
\hat{Y} = \hat{\mathcal{M}}\left(\vect{\Psi};\vect{\theta}_{\hat{\mathcal{M}}}\right) = \hat{\mathcal{M}}\left(\mathcal{H}\left(\vect{X};\vect{\theta}_{\mathcal{H}}\right);\vect{\theta}_{\hat{\mathcal{M}}}\right)\,, \label{Meta}
\end{equation}
\noindent where $\hat{\mathcal{M}}:\vect{\Psi}\in\mathbb{R}^{d_r}\mapsto\hat{Y}\in\mathbb{R}$ is a metamodel and $\vect{\theta}_{\hat{\mathcal{M}}}$ denotes parameters of the metamodel. The ultimate goal of dimensionality reduction in conjunction with metamodeling is to obtain an accurate estimate of the QoI, i.e., $P_f$ in Eq.\eqref{Pf}. Therefore, instead of training the dimensionality reduction and metamodel separately, it is desirable to train them jointly to minimize the error in failure probability estimation.

\section{Proposed method: AaS-hGP} \label{AaS-hGP}

\begin{figure}[h]
  \centering
  \includegraphics[scale=0.18]{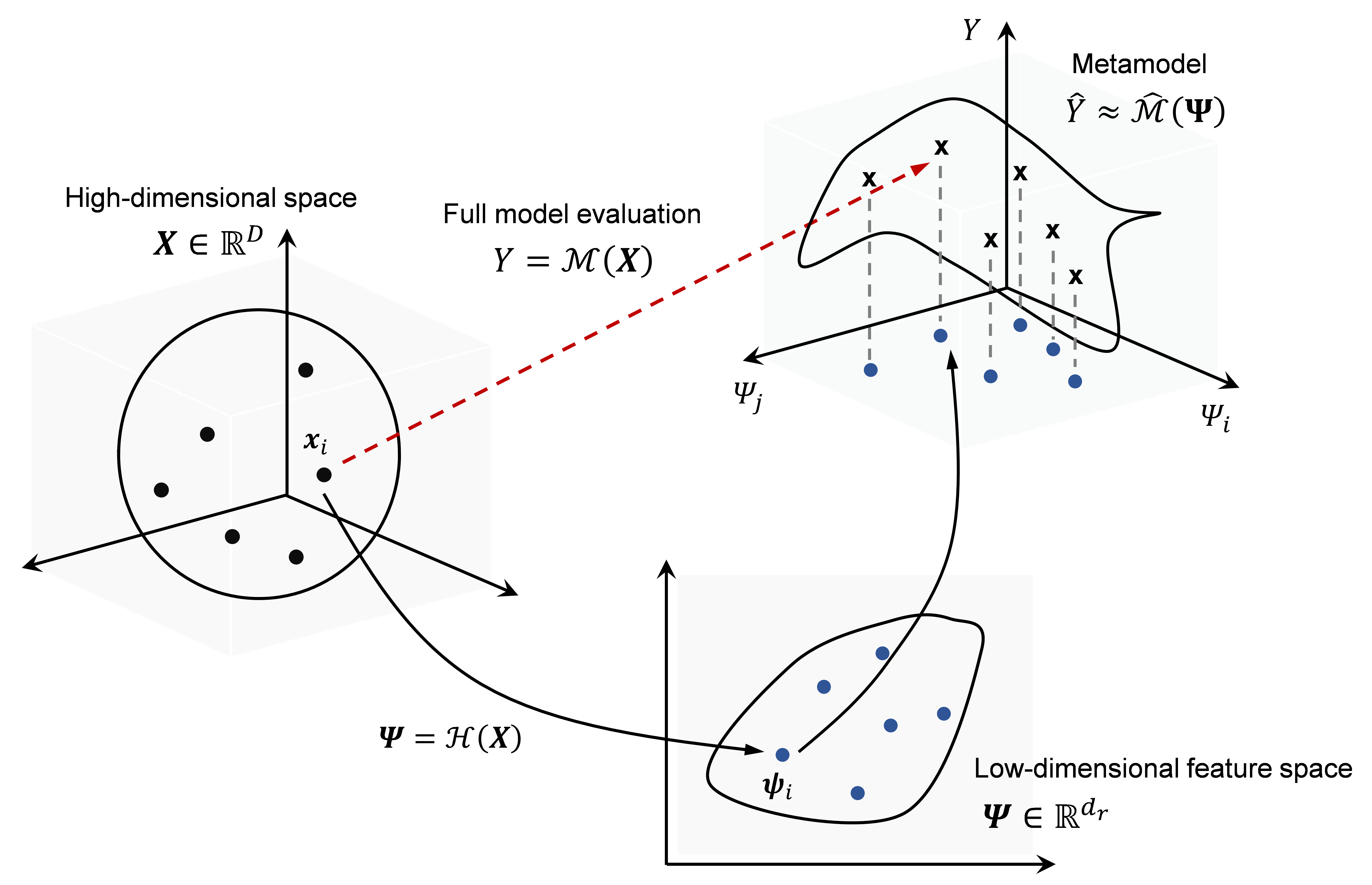}
  \caption{\textbf{Illustration of proposed metamodeling framework for high-dimensional reliability analysis}.} \label{Fig1}
\end{figure}

\noindent In this section, we address the reliability analysis problem featuring high-dimensionality and computationally expensive models by developing an adaptive active subspace-based metamodeling framework (Figure \ref{Fig1}). An active learning algorithm is also introduced to train the metamodel through an adaptive selection of training points achieving efficient convergence.

\subsection{Identification of low-dimensional features by active subspace}

\noindent An unsupervised dimensionality reduction does not use information from the computational model $\mathcal{M}$. This practice presumes that input $\vect{X}$ has a low-dimensional representation, which may not be appropriate for some reliability problems, e.g., problems with weakly correlated random variables (see \ref{App:A} for preliminary studies on various unsupervised dimensionality reduction schemes). As a supervised dimensionality reduction method, the active subspace leverages the tangent space of $\mathcal{M}$ to identify the low-dimensional features. Details of the active subspace scheme in the context of AaS-hGP are described as follows.

Assuming the model $\mathcal{M}$ in Eq.\eqref{ModelF} is continuous, differentiable, and square-integrable with respect to $f_{\vect{X}}(\vect{x})$, an active subspace is a subspace in which the model response exhibits significant variations. It is defined by a relatively small number $(\ll{D})$ of eigenvectors from a symmetric positive semi-definite ${D}\times{D}$ matrix $\vect{C}$ expressed as \cite{constantine2014active}
\begin{equation}
\vect{C}=\int{\nabla_{\vect{X}}{\mathcal{M}(\vect{x})}{\nabla_{\vect{X}}{\mathcal{M}(\vect{x})}}^{T}h_{\vect{X}}{(\vect{x})}}\,d\vect{x}=\mathbb{E}_h\left[\nabla_{\vect{X}}{\mathcal{M}(\vect{x})}{\nabla_{\vect{X}}{\mathcal{M}(\vect{x})}}^{T}\right]\, ,\label{C mat}
\end{equation}
\noindent where $\nabla_{\vect{X}}\mathcal{M}(\vect{x})$ denotes the gradient (column vector) of $\mathcal{M}(\cdot)$ with respect to $\vect{x}$ and $\mathbb{E}_h[\cdot]$ denotes the expectation with respect to a “learning kernel” $h_{\vect{X}}{(\vect{x})}$. In the proposed AaS-hGP framework, the learning kernel $h_{\vect{X}}{(\vect{x})}$ is controlled by the design of experiment (DoE) scheme. For example, if a global DoE using random samples of $f_{\vect{X}}{(\vect{x})}$ is adopted, then $h_{\vect{X}}{(\vect{x})}=f_{\vect{X}}{(\vect{x})}$. If an active learning algorithm guides the DoE, $h_{\vect{X}}{(\vect{x})}$ will evolve in the learning process. Notice that there is no need to prespecify a parametric form for $h_{\vect{X}}{(\vect{x})}$, as samples of $h_{\vect{X}}{(\vect{x})}$ will be generated from a prespecified DoE scheme. The matrix $\vect{C}$ is symmetric and positive semidefinite and thus admits the real eigenvalue decomposition:
\begin{equation}
\vect{C}=\vect{W}\vect{\Lambda}\vect{W}^T ,\,\,\,\,\,\vect{\Lambda}=diag(\lambda_{1},...,\lambda_{D})\,, \label{C decomp}
\end{equation}
\noindent where $\vect{\Lambda}$ is a diagonal matrix whose elements are the non-negative eigenvalues of $\vect{C}$ sorted in an descending order, i.e., $\lambda_{1}\geq...\geq\lambda_{D}\geq0$, and $\vect{W}$ is the ${D}\times{D}$ orthogonal matrix whose columns correspond to the eigenvectors of $\vect{C}$. One can partition the matrices of eigenvalues and eigenvectors as
\begin{equation}
\vect{\Lambda}=\left[
\begin{array} {cc}
    \vect{\Lambda}_r & \\
       & \vect{\Lambda}_s
 \end{array}
\right] ,\,\,\,\,\,\vect{W}=[\vect{W}_r \,\,\, \vect{W}_s]\,, \label{Lamb decomp}
\end{equation}
\noindent where $\vect{\Lambda}_r=diag(\lambda_1,...,\lambda_{d_r})$ denotes the first (and largest) $d_r$ eigenvalues and $\vect{W}_r$ denotes the first $d_r$ columns of $\vect{W}$. Assuming that a significant spectral gap between $\lambda_{d_r}$ and $\lambda_{d_r+1}$ exists, $\mathcal{M}(\vect{X})$ can be approximated as a low-dimensional representation using the subspace defined as
\begin{equation}
\vect{\Psi}=\vect{W}_{r}^{T}\vect{X}\, ,\label{AS map}
\end{equation}
\noindent where $\vect{\Psi}\in\mathbb{R}^{d_r}$ is the low-dimensional feature vector, termed \textit{active variable}, identified by the projection matrix $\vect{W}_r$. The projection matrix $\vect{W}_r$ defines the average change of $\mathcal{M}(\vect{X})$ in response to perturbations of $\vect{X}$ in the corresponding eigenvector direction, assuming that the variations in $\vect{W}_s$ directions are negligible. During the implementation, the projection operator is determined by a numerical $C$ obtained from gradient samples and the reduced dimensionality parameter $d_r$.

The reduced dimensionality $d_r$ can be a critical parameter significantly influencing the accuracy of the proposed method. Therefore, it is desirable to determine $d_r$ through the surrogate modeling, i.e., determine $d_r$ based on its measured contribution to the accuracy of surrogate model predictions. Consider $n$ training samples of $\vect{x}_i,i=1,...,n$, the corresponding $y_i=\mathcal{M}(\vect{x}_i)$, and $\nabla{y}_i=\nabla_{\vect{X}}\mathcal{M}(\vect{x}_i)$. A recursive algorithm (Algorithm \ref{alg:01}) is introduced as follows to find $d_r$ utilizing the following mean squared error measure:
\begin{equation}
\varepsilon_d=\sqrt{\frac{1}{n}\sum_{i=1}^{n}\left(y_i-\hat{\mathcal{M}}\left(\vect{\psi}_i; d_r, \vect{\theta}_{\hat{\mathcal{M}}}\right)\right)^2}\, ,\label{error_dr}
\end{equation}
where $\vect{\psi}_i$ denotes the feature space projection of $\vect{x}_i$, i.e., $\vect{\psi}_i=\vect{W}_{r}^{T}\vect{x}_i$; $\hat{\mathcal{M}}(\cdot)$ and 
$\vect{\theta}_{\hat{\mathcal{M}}}$ respectively denote metamodel predictions and the corresponding metamodeling parameters. Details of the metamodeling process will be introduced in the next section. The error measure in Eq.\eqref{error_dr} captures the contribution of $d_r$ to the surrogate modeling error. Starting from $d_r=1$, the AaS-hGP iteratively increases $d_r$ until $\varepsilon_d$ becomes smaller than a specified threshold $\varepsilon_d^t$. Note that the error measure is monitored at each learning step to find the optimal dimensionality given the current training dataset. It is also noted that given a dataset of $(\vect{x}_i,y_i,\nabla{y_i})$, the recursive procedure to find $d_r$ does not require additional evaluations of the model function $\mathcal{M}$.
\begin{breakablealgorithm}
\label{alg:01}
\caption{Adaptive rule to determine the optimal reduced dimensionality $d_r^{*}$ in AaS-hGP.}
\begin{description}

In each learning iteration, given $n$ training triples $[\vect{x}_i,y_i,\nabla{y_i}]^T$, $i=1,...,n$, and an average \\ derivative functional matrix $\vect{C}$ in Eq.\eqref{C mat}:
\\
$d_r\leftarrow{1}$;
\\
Identify the $d_r$-dimensional feature mapping described by Eq.\eqref{AS map};
\\
Predict surrogate model responses at $\vect{\psi}_i=\vect{W}_{r}^{T}\vect{x}_i$, i.e., $\hat{y}_i=\hat{\mathcal{M}}\left(\vect{\psi}_i; d_r, \vect{\theta}_{\hat{\mathcal{M}}}\right)$;
\\
Compute $\varepsilon_d$ in Eq.\eqref{error_dr};
\\
While $\varepsilon_d>\varepsilon_d^t$, do
\\
\,\,\,\,\,\,\,$d_r\leftarrow{d_r+1}$; 
\\
\,\,\,\,\,\,\,Identify the $d_r$-dimensional feature mapping described by Eq.\eqref{AS map};
\\
\,\,\,\,\,\,\,Predict surrogate model responses at $\vect{\psi}_i=\vect{W}_{r}^{T}\vect{x}_i$, i.e., $\hat{y}_i=\hat{\mathcal{M}}\left(\vect{\psi}_i; d_r, \vect{\theta}_{\hat{\mathcal{M}}}\right)$;
\\
\,\,\,\,\,\,\,Compute $\varepsilon_d$ in Eq.\eqref{error_dr};
\\
End
\\
$d_r^*\leftarrow{d_r}$; 

\end{description}
\end{breakablealgorithm}

\subsection{Response estimations in low-dimensional feature space using heteroscedastic Gaussian process}

\noindent A Gaussian process (GP)-based metamodel is employed to map low-dimensional feature vectors $\vect{\psi}_i$ into model predictions $\hat{y}_i$. Let us consider the active subspace mapping for a set of samples, $\mathcal{W}_\mathcal{D} = [\vect{W}_{r}^{T}\vect{x}_i,$ $ ... ,\vect{W}_{r}^{T}\vect{x}_n]^T$, and the corresponding model response observations $\mathcal{Y}_\mathcal{D} = \left[\mathcal{M}(\vect{x}_i),...,\mathcal{M}(\vect{x}_n)\right]^T$. The observations $(\mathcal{W}_\mathcal{D},\mathcal{Y}_\mathcal{D})$ in the reduced feature space may inevitably contain noises stemming from limited data and inherent errors of the feature space projection, further leading to the overfitting of surrogate models in the feature space \cite{nguyen2019ten}\cite{lataniotis2020extending}. Therefore, the possible variabilities stemming from the reduced feature representation should be incorporated into metamodeling. To this end, the heteroscedastic GP (hGP) model, which can incorporate input-dependent noises, is employed for model response predictions in the reduced feature space.

Introducing a Gaussian heteroscedastic noise $\varepsilon(\vect{\psi})$, the model response $y$ is represented by the following hGP model:
\begin{equation}
y=f(\vect{\psi})+\varepsilon(\vect{\psi})\,,\,\,\,\, \,\,\,\,\, \varepsilon(\vect{\psi})\sim
N(0,r(\vect{\psi}))\, ,\label{hGP_model}
\end{equation}
\noindent where $f(\vect{\psi})$ is a latent function for the response that is assumed to be a realization of a Gaussian process with mean $\mu_{f}(\vect{\psi})$ and covariance kernel $k_{f}(\vect{\psi},\vect{\psi}')$, and $r(\vect{\psi})=exp(g(\vect{\psi}))$ is the variance of the heteroscedastic noise parameterized to ensure positivity. $g(\vect{\psi})$ is a latent function introduced with a GP prior $g(\vect{\psi}) \sim GP(\mu_0,k_{g}(\vect{\psi},\vect{\psi}'))$ to handle the input-dependent noise. The hGP model consists of two latent functions $f(\vect{\psi})$ and $g(\vect{\psi})$ with (augmented) hyperparameters, i.e., $\vect{\theta}_{\hat{\mathcal{M}}} = \{\vect{\theta}_f,\vect{\theta}_g,\mu_{0} \}$, where $\vect{\theta}_f$ and $\vect{\theta}_g$ respectively denote the parameters for the kernel functions $k_{f}(\vect{\psi},\vect{\psi}')$ and $k_{g}(\vect{\psi},\vect{\psi}')$. Therefore, the analytical response prediction equations of a conventional GP model, described in \ref{App:B}, are not applicable to the hGP model.

The hyperparameters of the heteroscedastic GP model can be identified by the marginalized variational approximation scheme. Lázaro-Gredilla et al., (2013) introduced the following analytically tractable lower bound of the exact marginal likelihood using two sets of parameters – variational mean vector $\vect{m}$ and covariance matrix $\vect{V}$ \cite{lazaro2013retrieval}:
\begin{equation}
b_{MV}(\vect{m},\vect{V}) = \ln f_N(\vect{y}_\mathcal{D};\vect{0},\vect{K_f}+\vect{R}) - \frac{1}{4}tr(\vect{V}) - KL\left(f_N(\vect{g};\vect{m},\vect{V}) || f_N(\vect{g};\mu_{0}\vect{1},\vect{K_g})\right)    \, ,\label{hGP_MLE}
\end{equation}
\noindent where $f_N(\cdot)$ denotes the PDF of a multivariate Gaussian distribution, $tr(\cdot)$ is a trace operator, $KL(\cdot | \cdot)$ denotes the Kullback-Leibler divergence between two PDFs, $\vect{K_f}$ and $\vect{K_g}$ are respectively the covariance matrices of $f(\vect{\psi})$ and $g(\vect{\psi})$, and $\vect{R}$ is a diagonal matrix with elements $R_{i,i} = \exp(m_i - V_{i,i}/2), \, i=1,...,n$. Then, the hyperparameters can be estimated by maximizing this lower bound.

Given the hyperparameters, the predictive mean and variance functions of $y_*$ at a point $\vect{\psi}_*$ can be derived as \cite{lazaro2013retrieval}
\begin{equation}
\mu_{\hat{Y}}(\vect{\psi}_*) = \vect{k}_{\vect{f}_*}^T(\vect{K_f}+\vect{R})^{-1}\vect{y}_{\mathcal{D}}     \,, \label{hGP_mean}
\end{equation}
\begin{equation}
\sigma_{\hat{Y}}^2(\vect{\psi}_*) = \exp(\chi_* + \eta_*^2/2) + \gamma_*^2     \, ,\label{hGP_sigma}
\end{equation}
\noindent where $\chi_*$, $\eta_*^2$ and $\gamma_*^2$ denote the set of parameters described by kernel matrices of the two latent functions $f(\vect{\psi})$ and $g(\vect{\psi})$, i.e., $\chi_* = \vect{k}_{\vect{g}_*}^T(\vect{\Lambda}-\frac{1}{2}\vect{I})\vect{1}+\mu_0$, $\gamma_*^2 = k_{f_{**}} - \vect{k}_{\vect{f}_*}^T (\vect{K_f} + \vect{R})^{-1} \vect{k}_{\vect{f}_*}$, and $\eta_*^2 = k_{g_{**}}-\vect{k}_{\vect{g}_*}^T(\vect{K_g}+\vect{\Lambda}^{-1})^{-1}\vect{k}_{\vect{g}_*}$; $\vect{k}_{\vect{f}_*}$ and $\vect{k}_{\vect{g}_*}$ are vectors of covariance functions between prediction location $\vect{\psi}_*$ and $n$ observed points $\mathcal{W}_\mathcal{D}$ for latent functions $f(\vect{\psi})$ and $g(\vect{\psi})$, respectively; and $\vect{\Lambda}$ is a positive semi-definite diagonal matrix introduced to re-parametrize the variational parameters $\vect{m}$ and $\vect{V}$ in a reduced order.

\begin{figure}[h]
  \centering
  \includegraphics[scale=0.7] {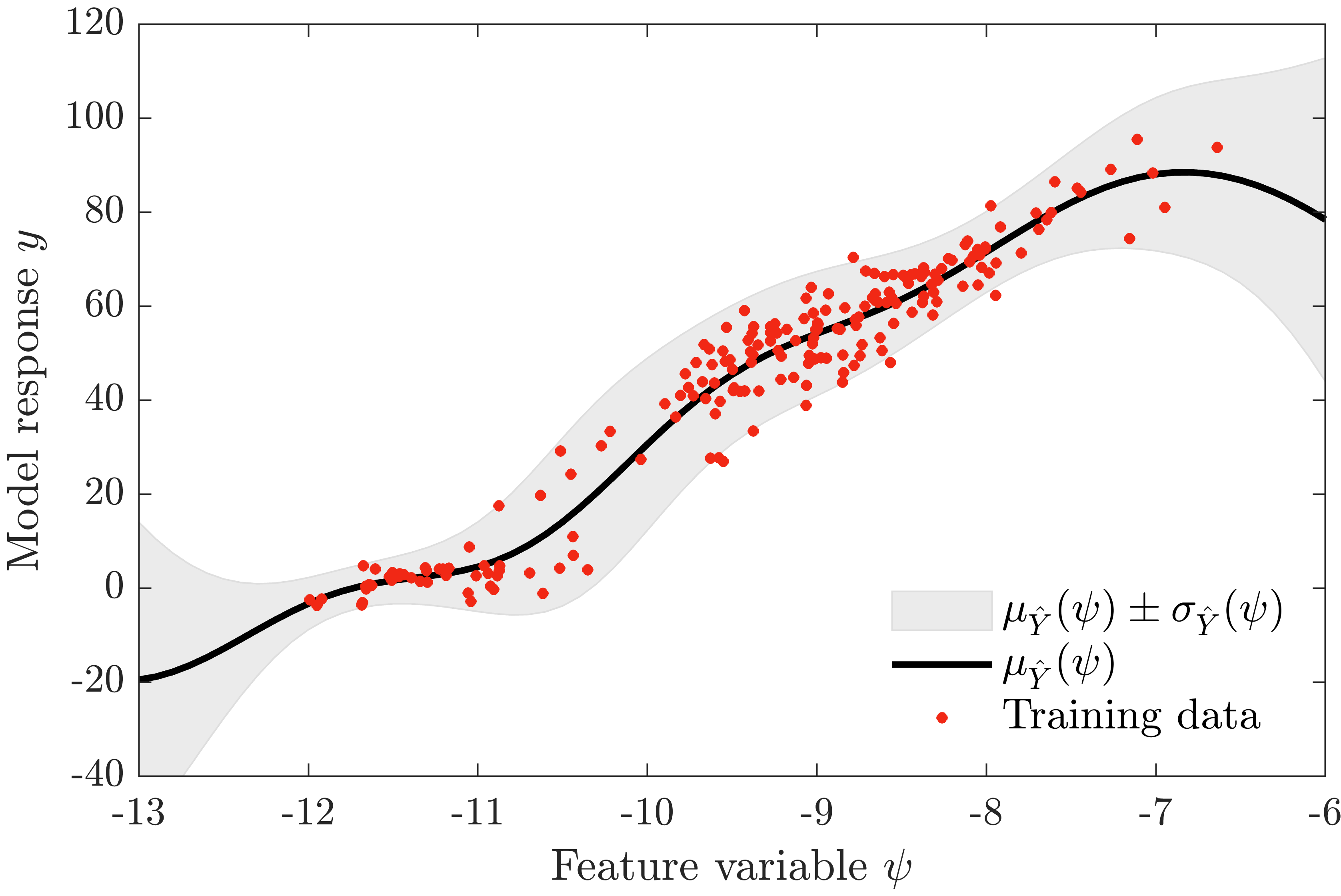}
  \caption{\textbf{Illustration of an hGP model and its predictions using a feature variable}. \emph{Due to imperfections of the feature space projection, the feature-response samples contain heteroscedastic noise. The hGP model has the desirable property to capture such heteroscedastic noise.}} \label{Fig2}
\end{figure}

The formulas of $\mu_{\hat{Y}}(\vect{\psi}_*)$ and $\sigma_{\hat{Y}}^2(\vect{\psi}_*)$ in Eqs. \eqref{hGP_mean} and \eqref{hGP_sigma} respectively serve as the mean and variance for the prediction of the model response $Y$, at a feature vector $\vect{\psi}=\vect{\psi}_*$ under the assumption of heteroscedastic noises. Note that both quantities can be utilized in an active learning scheme, which will be described in the next section. Figure \ref{Fig2} illustrates an example of the hGP predictions with projected observations (red dots) having heteroscedastic noises across the feature variable $\psi$.

\subsection{Adaptive training of metamodels by active learning}

\noindent The proposed AaS-hGP method employs an adaptive DoE strategy to train the surrogate model with limited training points \cite{blatman2011adaptive}\cite{kim2021clustering}. Since the computational target of AaS-hGP is a small failure probability, we need to enhance the fidelity of the metamodel in the critical region that significantly contributes to the failure probability. Consequently, the learning criterion should be designed to (1) guide the training process toward the vicinity of the limit-state surface, i.e., $\{\vect{x}\in\mathbb{R}^{D}:\mathcal{M}(\vect{x})=y_f\}$, and (2) ensure the training points are sparsely distributed to effectively capture properties of the model $\mathcal{M}$ in the critical region.

Thus, the learning criterion is developed such that the next training point $\vect{\psi}^*$ maximizes its distance to existing training points under the constraint that $\vect{\psi}^*$ is close to the limit-state surface, and the hGP prediction variance is large, i.e.,
\begin{equation}
\vect{\psi}^* = \mathop{\arg\max}_{\vect{\psi}\in\mathcal{W}_{c}} \min_{\vect{\psi}'\in\mathcal{W}_{\mathcal{D}}} \|\vect{\psi} - \vect{\psi}'\|     \, ,\label{AL1}
\end{equation}
\noindent where $\mathcal{W}_{\mathcal{D}}$ is a set of existing training points that are used to construct the current active subspace and metamodel, and $\mathcal{W}_{c}$ is a critical candidate training set defined by:
\begin{equation}
\mathcal{W}_{c} = \left\lbrace \vect{\psi}\in\mathcal{W}_{0}:\frac{|y_f - \hat{\mu}_{\hat{Y}}(\vect{\psi})|}{\hat{\sigma}_{\hat{Y}}(\vect{\psi})}\leq \varepsilon_{c} \right\rbrace\,, \label{AL2}
\end{equation}
\noindent where $\mathcal{W}_{0}$ is a candidate training set of random samples, mapped from the high-dimensional random samples of $\vect{X}$ generated by the PDF $f_{\vect{X}}{(\vect{x})}$; $\varepsilon_{c}$ is a cutoff value for learning ($\varepsilon_{c}=2.0$ is recommended); and $\hat{\mu}_{\hat{Y}}(\vect{\psi})$ and 
$\hat{\sigma}_{\hat{Y}}(\vect{\psi})$ are respectively the hGP-based mean and standard deviation of the prediction from Eqs. \eqref{hGP_mean} and \eqref{hGP_sigma} at the location $\vect{\psi}$. Eq.\eqref{AL1} and Eq.\eqref{AL2} embody an exploration-exploitation trade-off to choose the training sample located near the target failure surface and being most different from existing training points. After identifying the next training point $\vect{\psi}^*$, the corresponding high-dimensional input $\vect{x}^*$ can be identified, and the model response $y^*=\mathcal{M}(\vect{x}^*)$ and its gradients $\nabla{y}^*=\nabla_{\vect{X}}\mathcal{M}(\vect{x}^*)$ are evaluated to update the active subspace and the hGP model. It is important to note that the training point $\vect{\psi}^*$ is selected not by searching in $\mathbb{R}^{d_r}$ but by searching in a prespecified discretized set of training candidates with known $(\vect{x},\vect{\psi})$ pairings. Consequently, the mapping between $\vect{\psi}$ and $\vect{x}$ is pointwise one-to-one for all points from the discretized training candidate set. This property further suggests that AaS-hGP does not require the reconstruction mapping $\mathcal{H}^{-1}:\vect{\Psi}\in\mathbb{R}^{d_r}\mapsto{\hat{X}}\in\mathbb{R}^D$. 

During the active learning procedure, the failure probability $\hat{P}_f$ at each learning step is estimated using the trained hGP model and random samples, i.e.,
\begin{equation}
\hat{P}_f \cong \Prob{\hat{\mathcal{M}}(\vect{\Psi})\geq y_f} = \frac{\sum_{k=1}^{N}\mathbb{I}\left(\mu_{\hat{Y}}(\vect{\psi}_k) \geq y_f\right)}{N}  \, ,\label{PF_estim}
\end{equation}
\noindent where $\mathbb{I}\left(\mu_{\hat{Y}}(\vect{\psi}_k) \geq y_f\right)$ is a binary indicator function utilizing the response predictions in Eq.\eqref{hGP_mean}, which provides “1” if $\mu_{\hat{Y}}(\vect{\psi}_k) \geq y_f$ and “0” otherwise. The relative change of the failure probability estimate is used to construct a convergence criterion for AaS-hGP:
\begin{equation}
\epsilon_p^{(m)}=\left|\frac{\hat{P}_f^{(m)}-\hat{P}_f^{(m-1)}}{\hat{P}_f^{(m)}}\right|  \, ,\label{PF_conv}
\end{equation}
\noindent where $\hat{P}_f^{(m)}$ denotes the estimate of the failure probability at the $m$-th learning step. The AaS-hGP monitors whether $\epsilon_p^{(m)}$ in Eq.\eqref{PF_conv} at two successive steps becomes \textit{small} and \textit{stagnant} as the active learning proceeds. Specifically, the value at the $m$-th learning step, $\epsilon_1^{(m)}=\epsilon_p^{(m)}$, and the convergence “trend” defined as $\epsilon_2^{(m)}=|\epsilon_1^{(m-1)} - \epsilon_1^{(m)}|$ are calculated at each iteration. If both $\epsilon_1^{(m)}$ and $\epsilon_2^{(m)}$ become smaller than the specified tolerances $\epsilon_1^{tol}$ and $\epsilon_2^{tol}$, respectively, the active learning iteration is terminated.
\\
\\
\begin{figure}[h]
  \centering
  \includegraphics[scale=0.25] {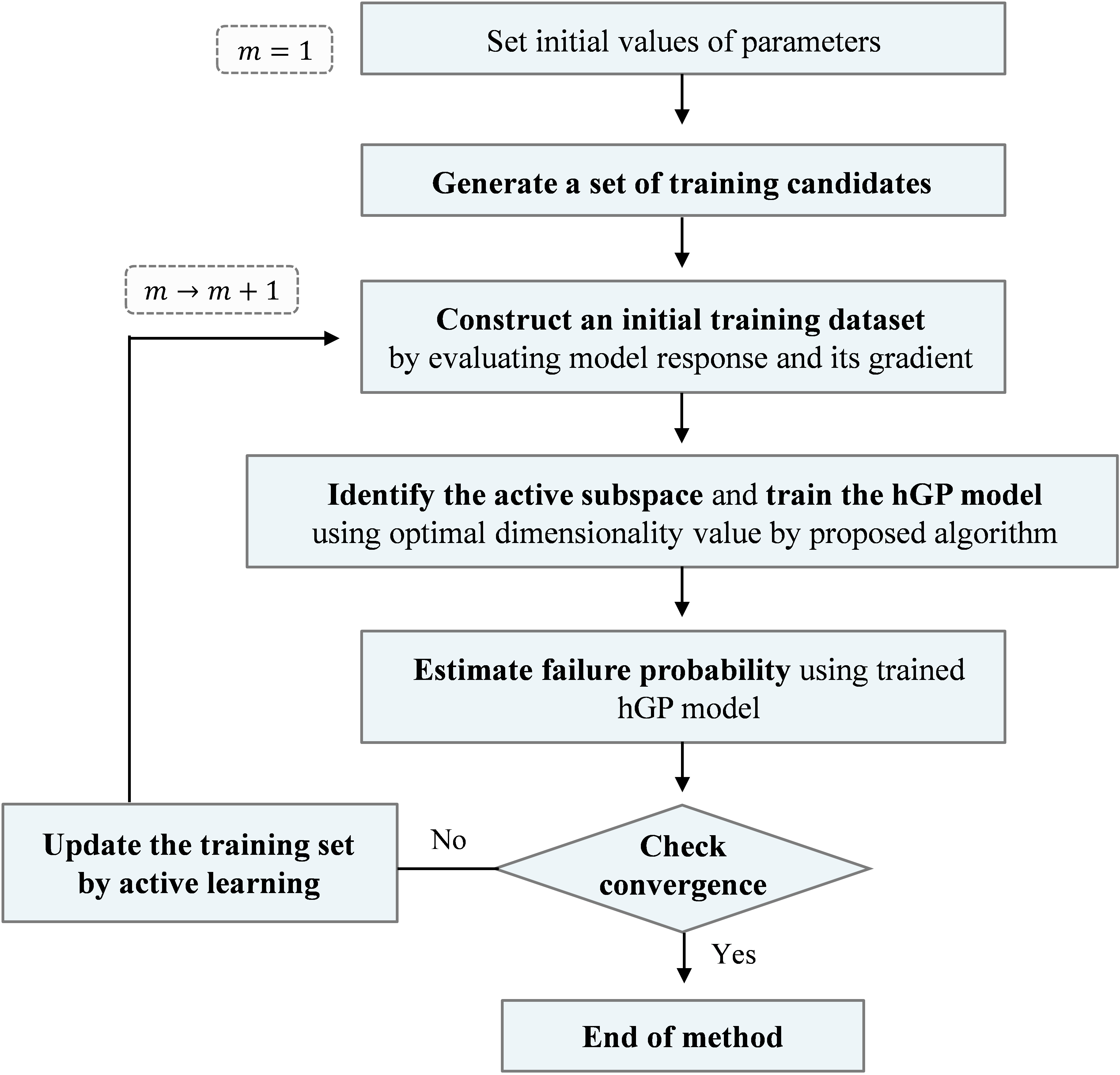}
  \caption{\textbf{Flowchart of AaS-hGP algorithm}.} \label{Fig3}
\end{figure}
\\
\subsection{Algorithm of AaS-hGP}  \label{Algorithmsection}

\noindent The basic procedure of the proposed AaS-hGP approach for high-dimensional reliability analysis, illustrated by the flowchart in Figure \ref{Fig3}, is summarized as follows:

\begin{description}
\item [Step 1: Initialization]
~\
\begin{itemize}
\item Set the convergence tolerances $\epsilon_1^{tol}$, $\epsilon_2^{tol}$ and $\varepsilon_d^{t}$, and determine the size of the initial training set, $n_0$ and that of the samples, $N$, $n_0 \ll N$.
\item Generate a set of training candidates $\mathcal{X}_0=\{\vect{x}_k, \, k=1,...,N\}$ from $f_{\vect{X}}(\vect{x})$. 
\item Generate an initial training set $\mathcal{X}_{\mathcal{D}} = \{\vect{x}_i, \, i=1,...,n_0\}$.
\item For each sample in $\mathcal{X}_\mathcal{D}$, evaluate the model response and its gradient to obtain sets $\mathcal{Y}_\mathcal{D}$ and $\nabla\mathcal{Y}_\mathcal{D}$. 
\end{itemize}

\item [Step 2: Identify the active subspace and train the hGP model]
~\
\begin{itemize}
\item Given $\mathcal{X}_\mathcal{D}$, $\mathcal{Y}_\mathcal{D}$ and $\nabla\mathcal{Y}_\mathcal{D}$, identify the active subspace mapping expressed by Eq.\eqref{AS map} and construct the hGP model in the active subspace using the optimal dimensionality value estimated from Algorithm \ref{alg:01}. For each of the trial dimensionality values in Algorithm \ref{alg:01}, an hGP model is trained by maximizing Eq.\eqref{hGP_MLE} using $\mathcal{W}_\mathcal{D}$ and $\mathcal{X}_\mathcal{D}$, where $\mathcal{W}_\mathcal{D}$ is obtained by projecting $\mathcal{X}_\mathcal{D}$ into the current active subspace.
\end{itemize}

\item [Step 3: Monte Carlo simulation by the hGP model]
~\
\begin{itemize}
\item Perform Monte Carlo simulation using the hGP model to obtain the failure probability.
\end{itemize}

\item [Step 4: Convergence check]
~\
\begin{itemize}
\item If the stopping criterion is met, terminate the algorithm and output the failure probability; else, proceed to \textbf{Step 5}.
\end{itemize}

\item [Step 5: Update the training set by active learning]
~\
\begin{itemize}
\item Project $\mathcal{X}_0$ into the current active subspace to obtain $\mathcal{W}_0$.
\item Solve Eq.\eqref{AL1} for the next training point $\vect{\psi}^*\in\mathcal{W}_0$ and find the corresponding $\vect{x}^*\in\mathcal{X}_0$. 
\item Add the new training point $\vect{x}^*$ to the training set $\mathcal{X}_\mathcal{D}$, i.e., $\mathcal{X}_\mathcal{D} \leftarrow \mathcal{X}_\mathcal{D} \cup \{\vect{x}^*\}$.
\item Compute $y^*=\mathcal{M}(\vect{x}^*)$ and $\nabla{y}^*=\nabla_{\vect{X}}\mathcal{M}(\vect{x}^*)$ at the new training point $\vect{x}^*$. 
\item Update $\mathcal{Y}_\mathcal{D}$ and $\nabla\mathcal{Y}_\mathcal{D}$ by $\mathcal{Y}_\mathcal{D} \leftarrow \mathcal{Y}_\mathcal{D} \cup \{y^*\}$ and $\nabla\mathcal{Y}_\mathcal{D} \leftarrow \nabla\mathcal{Y}_\mathcal{D} \cup \{\nabla{y}^*\}$.
\item Generate a new set of training candidates, $\mathcal{X}_0$.
\item Return to \textbf{Step 2}.
\end{itemize}
\end{description}

\section{Numerical examples} \label{Examples}

\noindent The proposed AaS-hGP method and its performance are demonstrated by three high-dimensional examples. The first example, a nonlinear mathematical model, is introduced to test the method's performance at various dimensions. The following two examples present a space truss structure and steel lattice transmission tower to further investigate the applicability of the proposed method to realistic structural reliability problems.

\subsection{Example 1: Nonlinear mathematical function}

\noindent Consider a nonlinear mathematical function formulated as \cite{marrel2008efficient}\cite{navaneeth2022surrogate}
\begin{equation}
Y = \mathcal{M}(\vect{X}) = \prod_{j=1}^{D}\mathcal{M}_j(X_j)\,,\,\, \,\,\, \mathcal{M}_j(X_j) = \frac{4X_j-2+\lambda_{j}}{1+\lambda_{j}}   \, ,\label{Examp1}
\end{equation}
\noindent where $X_j, \, j=1,...,D$ are basic random variables uniformly distributed over the range of [0, 1], $D$ is the dimension, and $\lambda_{j}$ is the non-negative model parameter that characterizes the influence of an input random variable $X_{j}$ on the output $Y$. Lower values of $\lambda_{j}$  indicate a significant first-order effect of $X_{j}$. Thus, the model function in Eq.\eqref{Examp1} is desirable to test whether the proposed scheme identifies the proper low-dimensional subspace. The parameter values are set to $\lambda_{j}=1$ if $j=1,...,4$ and $\lambda_{j}=500$ otherwise. Therefore, the effective dimension is four under this parametric setting. Note that, in general, the conventional unsupervised dimensionality reduction techniques, e.g., PCA and diffusion maps, cannot solve the current problem effectively since all random variables are independent, i.e., the probability space of $\vect{X}$ lacks a low-dimensional structure. The failure event is defined as the model output exceeding a prescribed threshold of $y_f=0.65$. Four dimensionality values, $D$=30,50,70, and 100 are investigated.

\begin{figure}[H]
  \centering
  \includegraphics[scale=0.55] {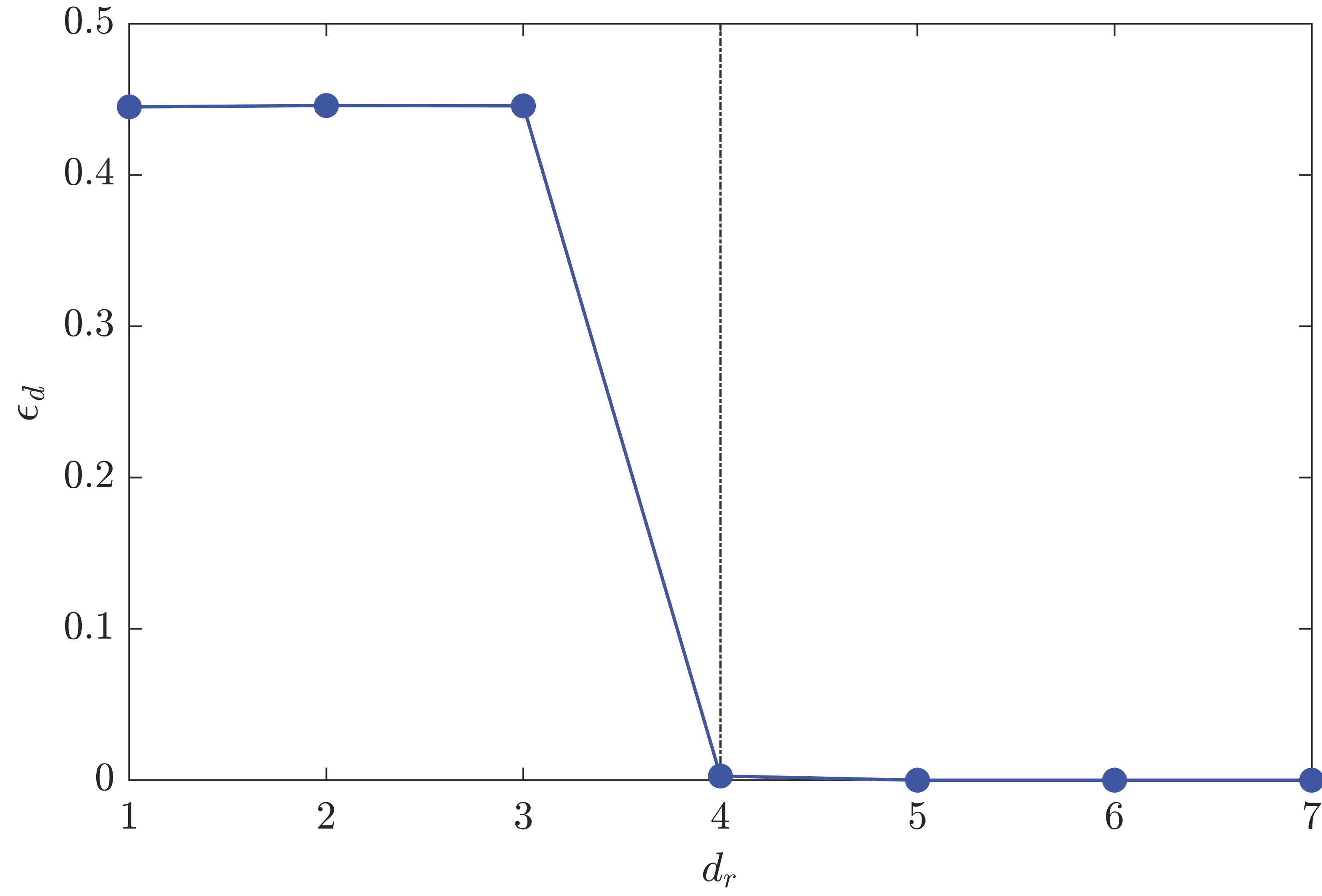}
  \caption{\textbf{Errors of surrogate modeling as a function of reduced dimensionality}. \emph{This figure is a byproduct of Algorithm 1. It suggests that the algorithm correctly identifies the effective dimensionality as $d_r=4$.}} \label{Fig4}
\end{figure}

\begin{figure}[H]
  \centering
  \includegraphics[scale=0.49] {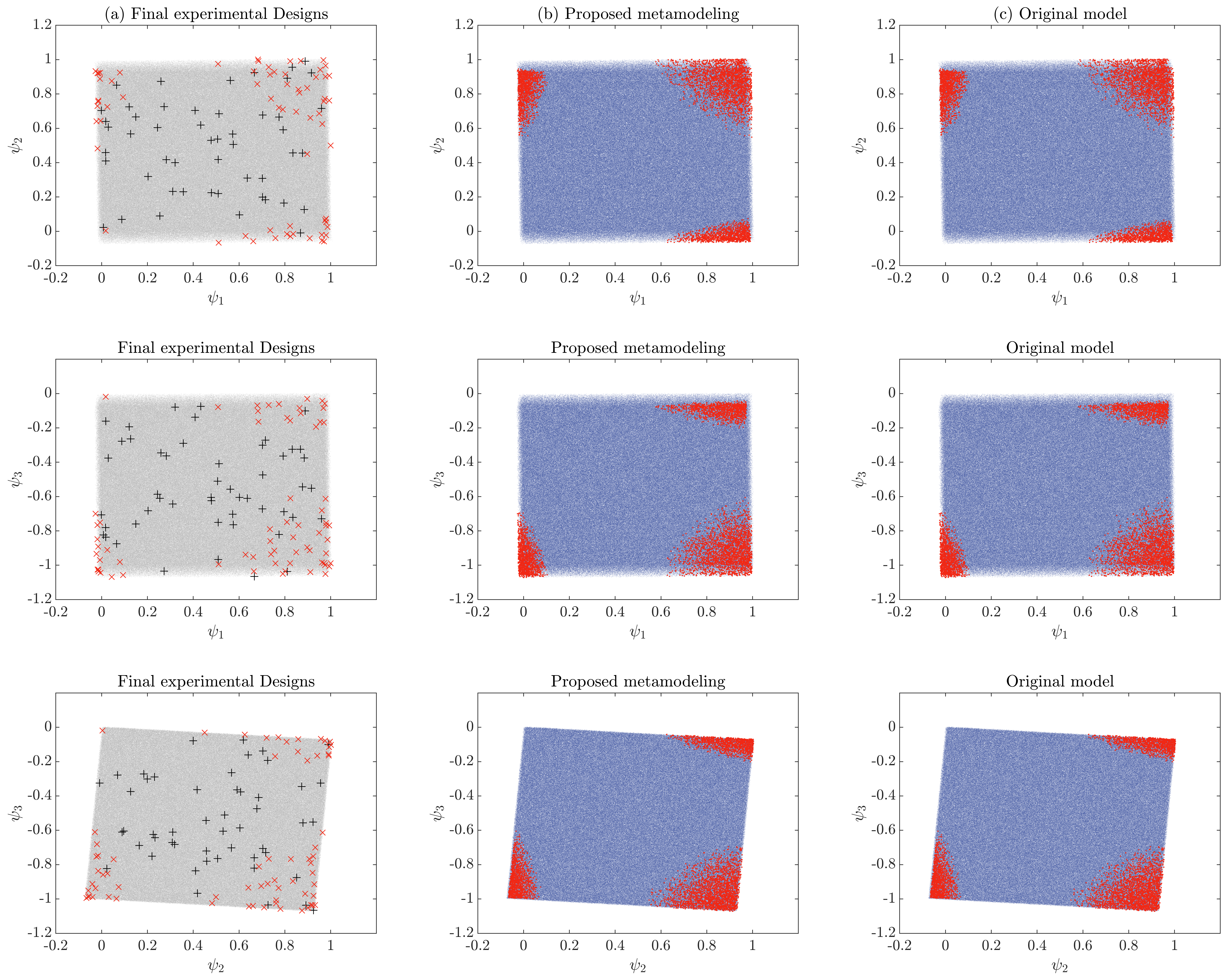}
  \caption{\textbf{Comparison of model responses represented in various planes of the reduced feature space at the final learning stage for Example 1 $(D=50)$: (a) final experimental designs, (b) predictions by the proposed method, and (c) original model responses}.} \label{Fig5}
\end{figure}

Following the procedure described in Section \ref{AaS-hGP}, the AaS-hGP method is applied with 50 initial DoEs. The convergence tolerances $\epsilon_1^{tol}$ and $\epsilon_2^{tol}$ are set to 0.001. Figure \ref{Fig4} illustrates the change of the mean square error in Eq.\eqref{error_dr} as the active subspace dimension varies, which is a byproduct of Algorithm \ref{alg:01} The result shows that $d_r=4$ is sufficient to achieve a small prediction error, and thus the proposed scheme determines a suitable dimensionality reduction consistent with the parameter values set in a priori. Figure \ref{Fig5} shows all training data points explored up to the final step of the proposed method, which are projected into various planes of the reduced feature space, i.e., $( \psi_1,\psi_2 )$, $( \psi_1,\psi_3 )$, and $( \psi_2,\psi_3 )$ planes. In the first column of Figure \ref{Fig5}, the gray circles denote samples from the candidate training set mapped from the original sample space. The black-plus markers denote the initial training points used to construct the initial metamodel. The red-cross markers are the training points adaptively added during the active learning phase. The second and third columns of Figure \ref{Fig5} compare the metamodel predictions with the original model responses in the reduced feature space; the blue-circle and red-cross markers denote samples identified as located in the safe and failure regions, respectively. The result confirms that the metamodel predictions are consistent with the actual model responses, and the proposed method captures all failure patterns in the reduced feature space.

The reliability analysis results are presented in Table \ref{Tab_ex1}, which shows failure probability estimates obtained by 20 independent runs and the average number of model evaluations. The results are compared with those by direct MCS, First-order reliability methods (FORM) \cite{der2022structural}, and AK-MCS method \cite{echard2011ak}. For comparison, the “non-adaptive” results obtained by the active subspace-based metamodel predictions based on randomly selected (global) DoE points (using only Steps 2-3 of the AaS-hGP algorithm in Section \ref{Algorithmsection}) are also presented. The direct MCS solution $\hat{P}_{f,MCS}$ is considered the reference solution. The accuracy and efficiency of the reliability methods are compared in terms of $N_s$, i.e., the number of model function evaluations, the number of gradient evaluations $N_g$, the failure probability estimate $\hat{P}_f$, the generalized reliability index $\hat{\beta}_g = -\Phi^{-1}(\hat{P}_f)$, and the relative error with respect to the reference solution, i.e., $\varepsilon_p=|\hat{P}_{f} - \hat{P}_{f,MCS}|/{\hat{P}_{f,MCS}}$. For AaS-hGP, $N_s$ is the sum of the number of initial training points ($n_0$) and that of model evaluations during the active-learning process. As shown in Table \ref{Tab_ex1}, FORM analysis yields a large error due to linear approximations. AK-MCS requires many model simulations (more than 3,000) and still provides inaccurate results. It is noted that AK-MCS analysis is infeasible for $D>50$ since a large number of input variables makes active learning and metamodeling ineffective. It is observed that the proposed method produces accurate estimates of failure probabilities in high-dimensional problems by a small number of model function and gradient calls. In contrast, the “non-adaptive” method yields a larger error even by using more training points.

\begin{figure}[H]
  \centering
  \includegraphics[scale=0.65] {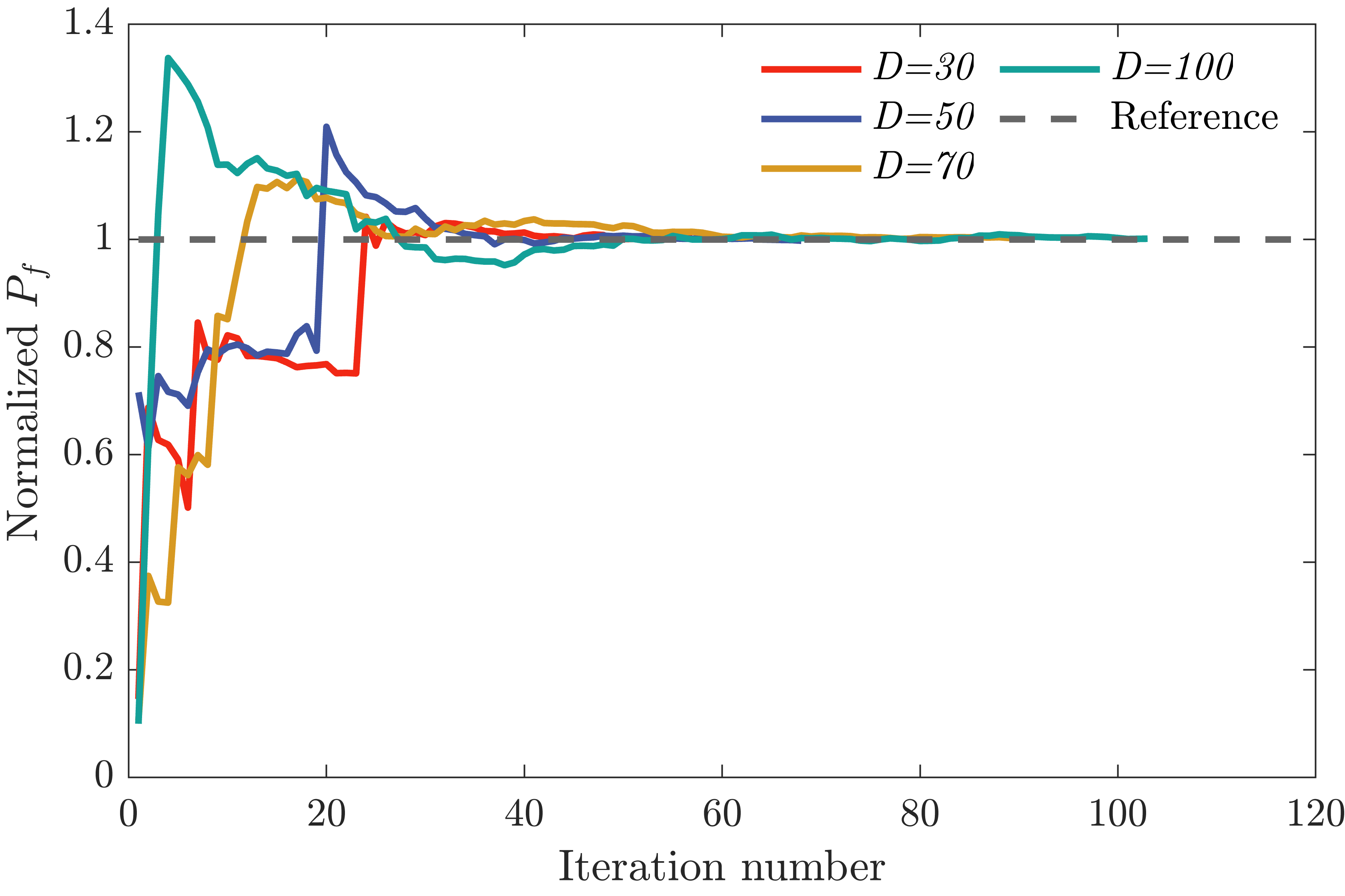}
  \caption{\textbf{Convergence histories of the proposed method for Example 1}.} \label{Fig6}
\end{figure}

\begin{figure}[H]
  \centering
  \includegraphics[scale=0.6] {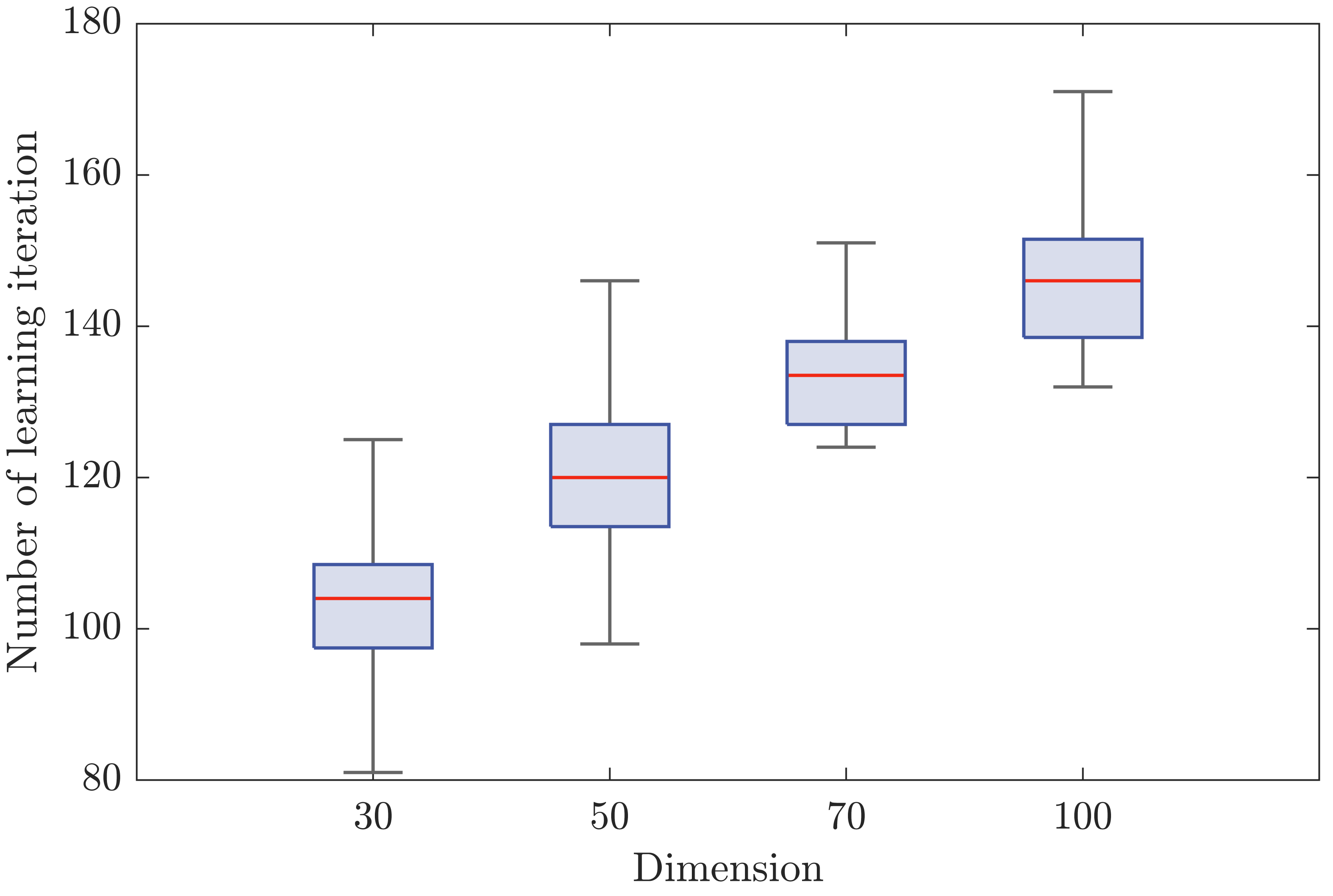}
  \caption{\textbf{Number of required learning iterations at various dimensions for Example 1}.} \label{Fig7}
\end{figure}

Figure \ref{Fig6} shows the convergence histories of the AaS-hGP analysis for four different dimensions. The estimated failure probability is normalized in the plot, i.e., it is divided by $\hat{P}_{f,MCS}$. In each case, the failure probability estimate approaches the reference solution with more iterations (training points). The box plot of the number of required learning iterations for different dimensions is shown in Figure \ref{Fig7}, demonstrating that AaS-hGP can efficiently handle high-dimensional reliability problems using a limited computational budget.

\begin{table}[H]
  \caption{\textbf{Performance of AaS-hGP compared with other reliability analysis methods for Example 1}.}
  \label{Tab_ex1}
  \centering
  \begin{center}
  \begin{tabular}{c c c c c c c}
    \toprule
    Dimension & Methods & $\hat{P}_{f}$ & $N_s$ & $N_g$ & $\hat{\beta}_{g}$ & $\varepsilon_p(\%)$\\
    \midrule
    \multirow{6}{4em}{$D=30$} & MCS & $5.77\times10^{-3}$ & $10^6$ & - & 2.53 & - \\
    & FORM & $3.86\times10^{-4}$ & 12 & 12 & 3.36 & 93.31 \\
    & AK-MCS & $4.00\times10^{-5}$ & $>3,000$ & - & 3.94 & 99.30 \\
    & \multirow{2}{8em}{\centering{ AaS-hGP\\(w/ global DoEs)}} & \multirow{2}{6em}{\centering{$5.81\times10^{-3}$}} & \multirow{2}{3em}{\centering{300}} & \multirow{2}{3em}{\centering{300}} & \multirow{2}{3em}{\centering{2.52}} & \multirow{2}{3em}{\centering{0.69}} \\
    &  &  &  &  & &  \\ 
    & AaS-hGP & $5.80\times10^{-3}$ & 104 & 104 & 2.52 & 0.52 \\
    \midrule
    \multirow{6}{4em}{$D=50$} & MCS & $4.96\times10^{-3}$ & $10^6$ & - & 2.58 & - \\
    & FORM & $2.18\times10^{-5}$ & 16 & 16 & 4.09 & 99.56 \\
    & AK-MCS & $8.20\times10^{-5}$ & $>3,000$ & - & 3.77 & 98.35 \\
    & \multirow{2}{8em}{\centering{ AaS-hGP\\(w/ global DoEs)}} & \multirow{2}{6em}{\centering{$4.51\times10^{-3}$}} & \multirow{2}{3em}{\centering{450}} & \multirow{2}{3em}{\centering{450}} & \multirow{2}{3em}{\centering{2.61}} & \multirow{2}{3em}{\centering{9.07}} \\
    &  &  &  &  & &  \\ 
    & AaS-hGP & $4.95\times10^{-3}$ & 123 & 123 & 2.58 & 0.21 \\
    \midrule
    \multirow{6}{4em}{$D=100$} & MCS & $3.33\times10^{-3}$ & $10^6$ & - & 2.71 & - \\
    & FORM & $1.78\times10^{-8}$ & 26 & 26 & 5.51 & 99.99 \\
    & AK-MCS & - & - & - & - & - \\
    & \multirow{2}{8em}{\centering{ AaS-hGP\\(w/ global DoEs)}} & \multirow{2}{6em}{\centering{$3.06\times10^{-3}$}} & \multirow{2}{3em}{\centering{700}} & \multirow{2}{3em}{\centering{700}} & \multirow{2}{3em}{\centering{2.74}} & \multirow{2}{3em}{\centering{8.11}} \\
    &  &  &  &  & &  \\ 
    & AaS-hGP & $3.33\times10^{-3}$ & 147 & 147 & 2.71 & 0.15 \\
    \bottomrule
  \end{tabular}
  \end{center}
\end{table}

\subsection{Example 2: Space truss structure}

\begin{figure}[H]
  \centering
  \includegraphics[scale=0.21] {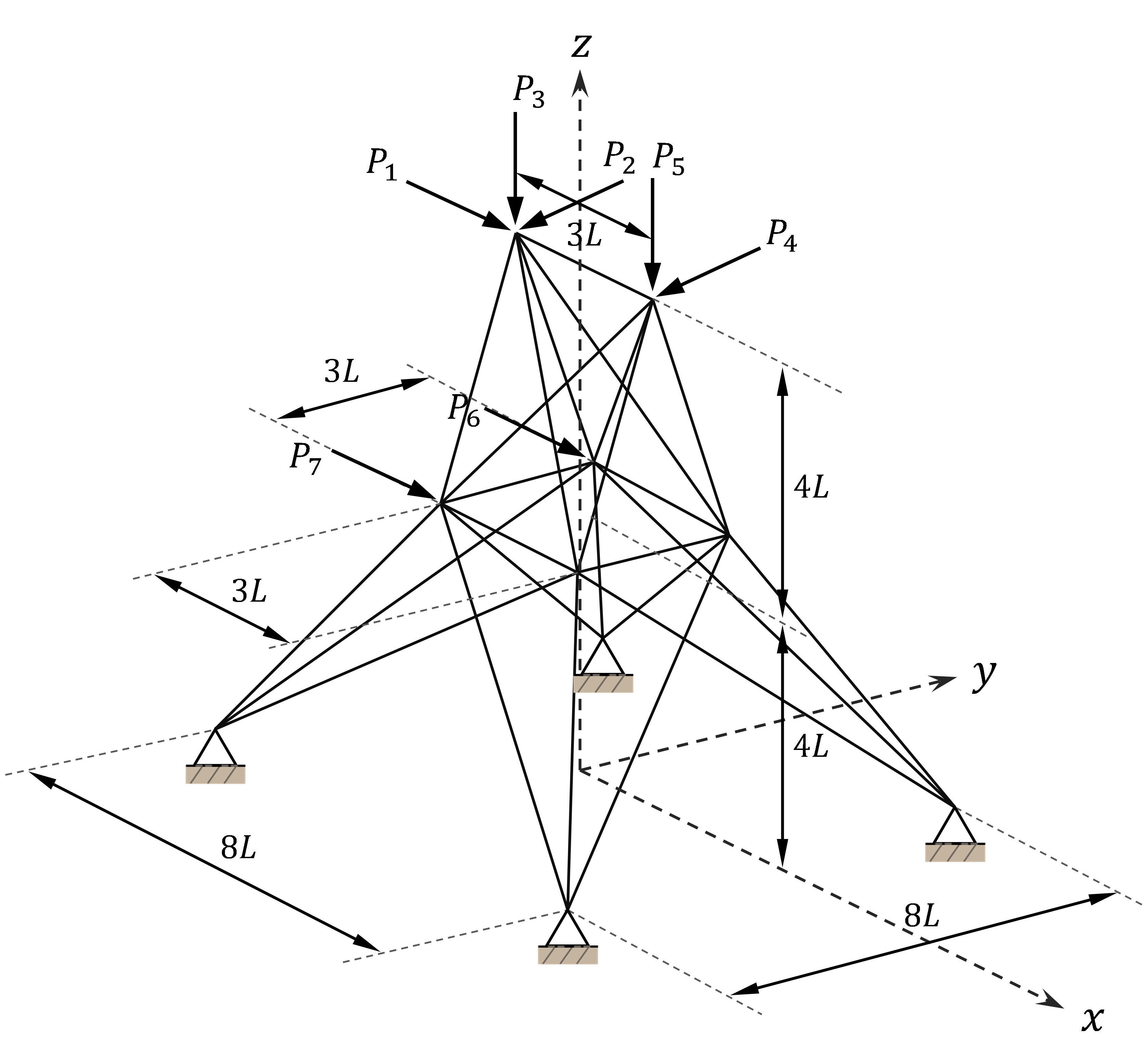}
  \caption{\textbf{A space truss structure}.} \label{Fig8}
\end{figure}

\noindent As an engineering application, this example deals with a space truss structure illustrated in Figure \ref{Fig8} \cite{rahami2008sizing}. The truss consists of twenty-five elements with ten nodes and is subjected to five horizontal $(P_1,P_2,P_4,P_6$ and $P_7)$ and two vertical $(P_3$ and $P_5)$ loads. The model response function is defined as the maximum displacement of the system, which can be expressed as
\begin{equation}
Y = \mathcal{M}(\vect{X}) = \max{(u_h(\vect{X}),u_v(\vect{X}))}   \, ,\label{Examp2}
\end{equation}
\noindent where $u_h$ and $u_v$ denote the peak displacements in horizontal and vertical directions at the top of the structure. The random vector $\vect{X}$ includes fifty-seven independent random variables, i.e., $\vect{X}=[X_1,X_2,...,X_{57}]$, that affect structural responses. These random variables are associated with external forces, cross-sectional areas, and modulus of elasticity. Table \ref{Tab_ex2} summarizes the distribution models and parameters of all random variables. Structural failure is defined as the maximum response exceeds a prescribed threshold of $y_f=0.45$ \textit{in}.

\begin{table}[H]
  \caption{\textbf{Distribution models and parameters of the random variables in Example 2}.}
  \label{Tab_ex2}
  \centering
  \begin{tabular}{c c c c c}
    \toprule
    Variable number & Random variables & Distribution & Mean & c.o.v. \\
    \midrule
    1 & $P_1 \, (N)$ & Lognormal & 1000 & 0.1 \\
    2-5 & $P_2-P_5 \, (N)$ & Lognormal & 10000 & 0.05 \\
    6 & $P_6 \, (N)$ & Lognormal & 600 & 0.1 \\
    7 & $P_7 \, (N)$ & Lognormal & 500 & 0.1 \\
    8-32 & $E_1-E_{25} \, (m^2)$ & Lognormal & $10^7$ & 0.05 \\
    33 & $A_1 \, (m^2)$ & Gaussian & 0.4 & 0.1 \\
    34-37 & $A_2-A_5 \, (m^2)$ & Gaussian & 0.1 & 0.1 \\
    38-41 & $A_6-A_9 \, (m^2)$ & Gaussian & 3.4 & 0.1 \\
    42-43 & $A_{10}-A_{11} \, (m^2)$ & Gaussian & 0.4 & 0.1 \\
    44-45 & $A_{12}-A_{13} \, (m^2)$ & Gaussian & 1.3 & 0.1 \\
    46-49 & $A_{14}-A_{17} \, (m^2)$ & Gaussian & 0.9 & 0.1 \\
    50-53 & $A_{18}-A_{21} \, (m^2)$ & Gaussian & 1.0 & 0.1 \\
    54-57 & $A_{22}-A_{25} \, (m^2)$ & Gaussian & 3.4 & 0.1 \\
    \bottomrule
  \end{tabular}
\end{table}

Table \ref{Tab_ex2_2} presents the estimated failure probabilities and the numbers of model function and gradient evaluations. The MCS results are obtained using $10^6$ structural simulations. It is noted that the AK-MCS method cannot provide a reliable estimate even with a large number of training points. It is noted that AaS-hGP produces accurate results consistent with the reference solution $\hat{P}_{f,MCS}$ at a significantly reduced computational cost compared to other reliability approaches.

\begin{figure}[H]
  \centering
  \includegraphics[scale=0.6] {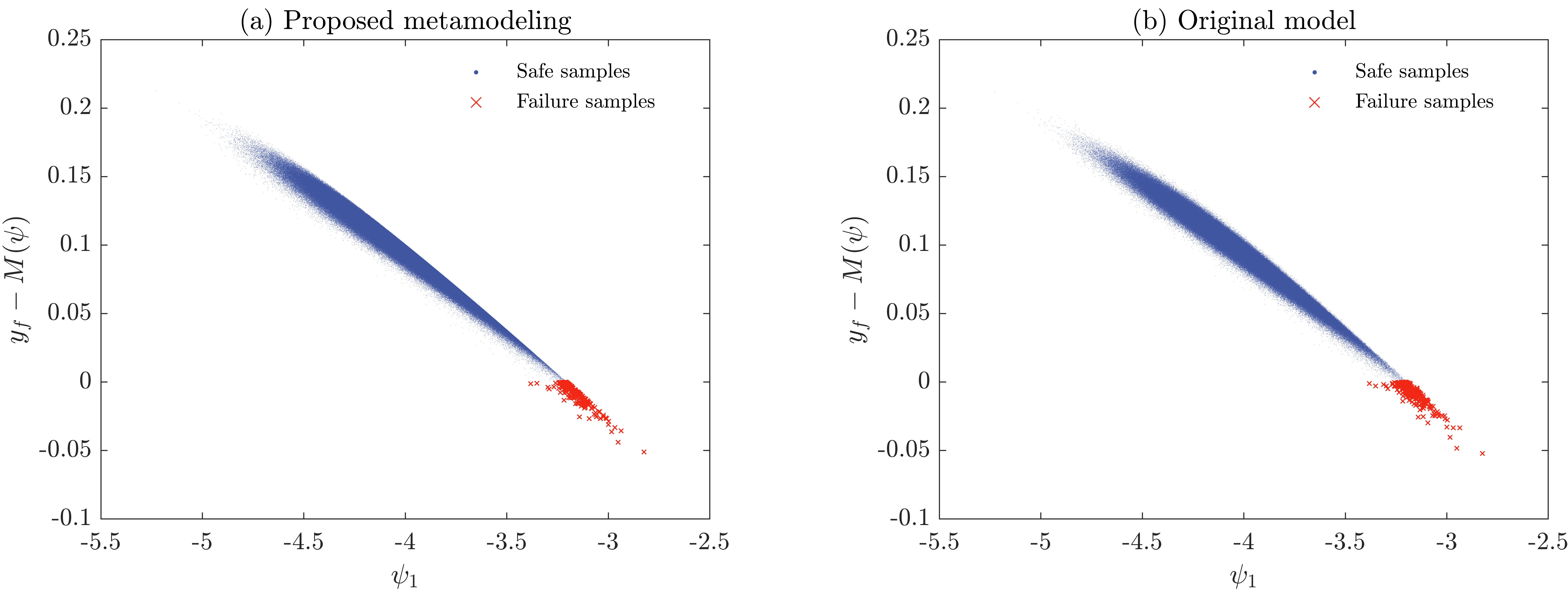}
  \caption{\textbf{Model responses against the dominant feature at the final learning stage for Example 2: (a) predictions by the proposed method and (b) original model responses}.} \label{Fig9}
\end{figure}

\begin{figure}[H]
  \centering
  \includegraphics[scale=0.6] {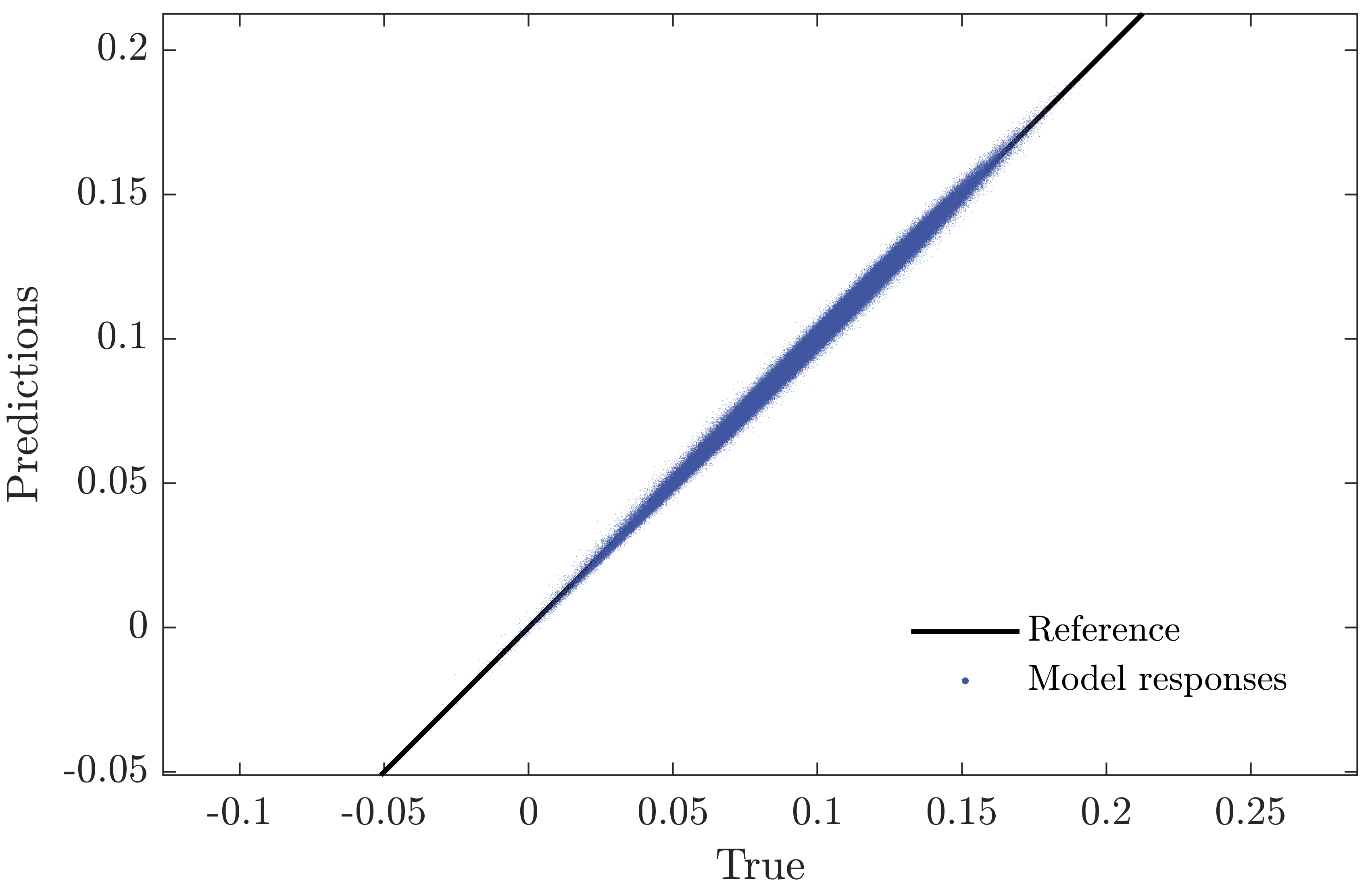}
  \caption{\textbf{Scatter plot of the model responses obtained from the proposed metamodel predictions and MCS for Example 2}.} \label{Fig10}
\end{figure}

Figure \ref{Fig9} compares the metamodel predictions with the original model responses, plotted against the first, i.e., the most dominant, feature $\psi_1$ of the active subspace. The blue-circle and red-cross markers denote samples in the safe and failure domains, respectively. The result confirms that the proposed metamodeling approach produces predictions consistent with the actual model responses. Figure \ref{Fig10} presents scatter plots of the model response $y$ against the surrogate prediction $\hat{y}$. Ideally, the scatter plot should yield a line. The results again confirm the accuracy of the proposed approach for high-dimensional reliability analysis. 

\begin{figure}[H]
  \centering
  \includegraphics[scale=0.6] {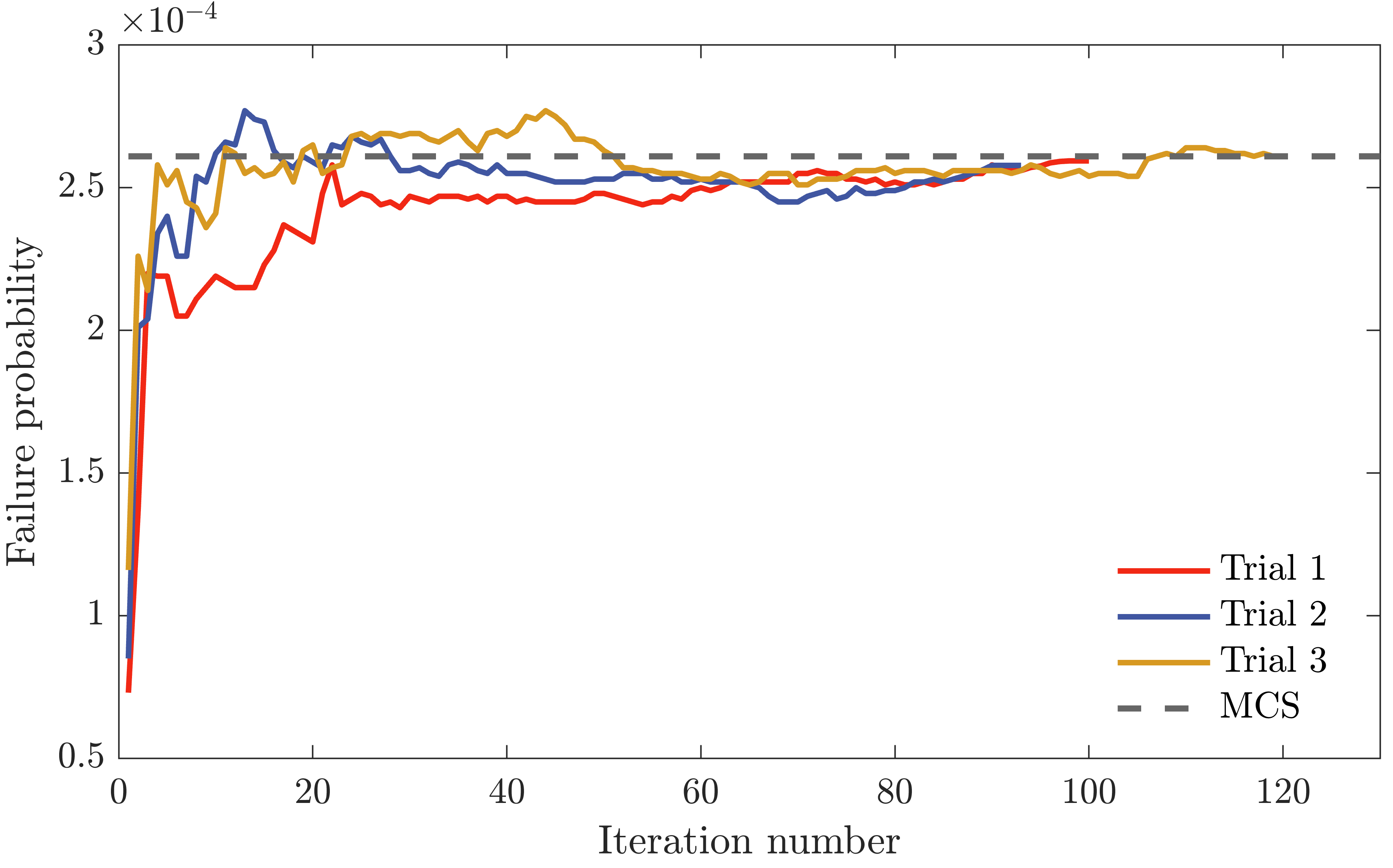}
  \caption{\textbf{Convergence histories of the proposed method for Example 2}.} \label{Fig11}
\end{figure}

The convergence histories of AaS-hGP analysis starting from different initial training data points are presented in Figure \ref{Fig11}. In each case, the failure probability estimate gets close to the reference solution after 100 learning iterations despite the randomness in the initial training dataset. The convergence histories confirm that the active learning scheme of AaS-hGP works effectively to ensure the method is robust against possible variations in the initial training points.

\begin{table}[H]
  \caption{\textbf{Performance of AaS-hGP compared with other reliability analysis methods for the space truss example}.}
  \label{Tab_ex2_2}
  \centering
  \begin{tabular}{c c c c c c}
    \toprule
    Methods & $\hat{P}_{f}$ & $N_s$ & $N_g$ & $\hat{\beta}_{g}$ & $\varepsilon_p(\%)$\\
    \midrule
    MCS & $2.61\times10^{-4}$ & $10^6$ & - & 3.47 & - \\
    FORM & $1.08\times10^{-7}$ & 43 & 43 & 5.18 & 99.96 \\
    AK-MCS & $2.86\times10^{-5}$ & $>3,000$ & - & 4.02 & 89.05 \\
    \multirow{2}{8em}{\centering{ AaS-hGP\\(w/ global DoEs)}} & \multirow{2}{6em}{\centering{$1.72\times10^{-4}$}} & \multirow{2}{3em}{\centering{500}} & \multirow{2}{3em}{\centering{500}} & \multirow{2}{3em}{\centering{3.58}} & \multirow{2}{3em}{\centering{34.10}} \\
    &  &  &  &  &  \\ 
    AaS-hGP & $2.63\times10^{-4}$ & 153 & 153 & 3.47 & 0.76 \\
    \bottomrule
  \end{tabular}
\end{table}

\subsection{Example 3: Steel lattice transmission tower}

\begin{figure}[H]
  \centering
  \includegraphics[scale=0.21] {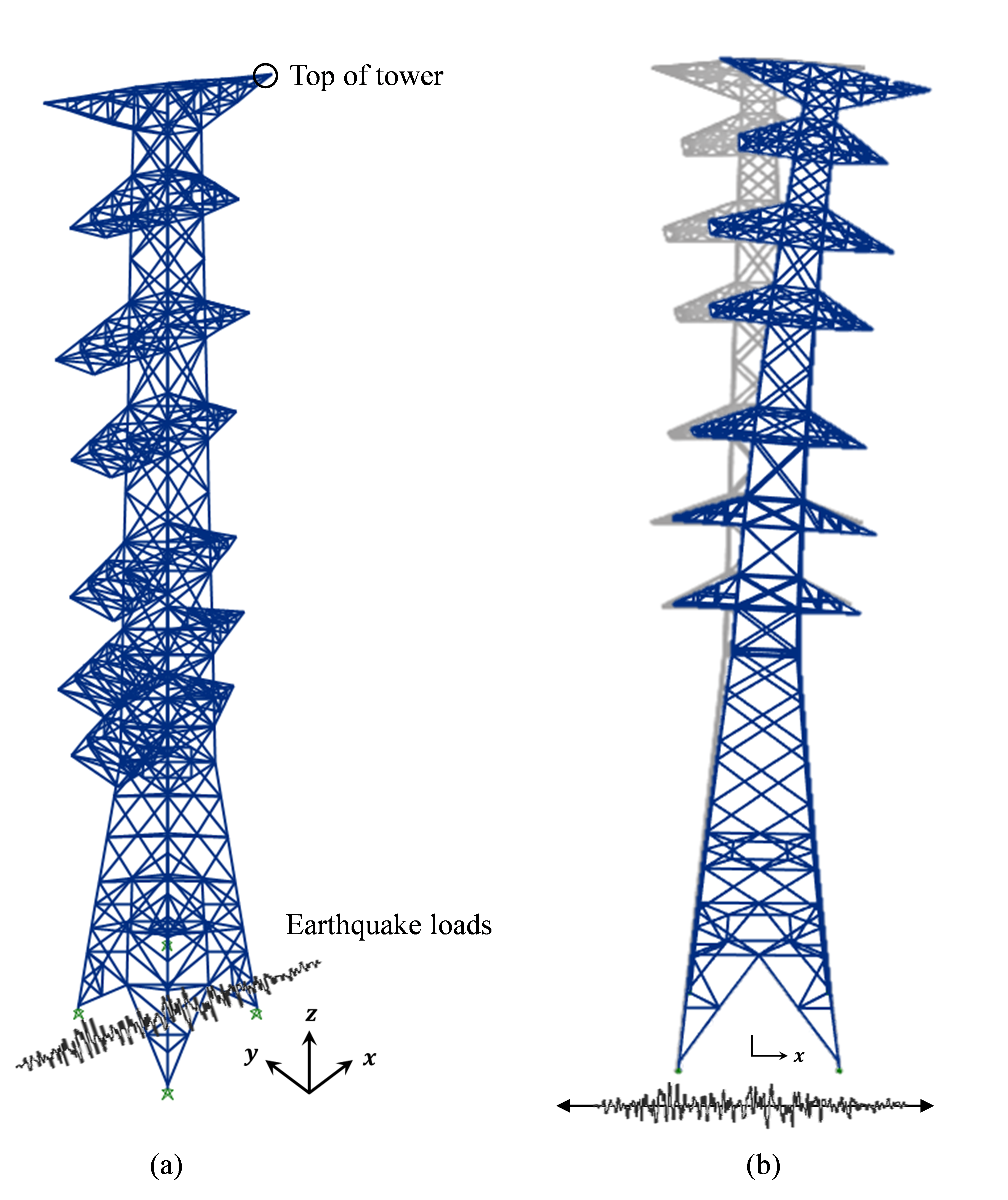}
  \caption{\textbf{Finite element model of the transmission tower: (a) perspective view and (b) front view}.} \label{Fig12}
\end{figure}

\noindent This example examines a transmission tower subjected to seismic loads. Figure \ref{Fig12} shows the finite element model of the transmission tower, created by the SAP2000 software to perform the nonlinear time-history analyses \cite{park2016seismic}. The tower's height is 86.6 \textit{m}, and each section consists of several continuous panels with cross-arms. The foundation is assumed to be rigid, i.e., the model is fixed at the base. The real ground motions of the El Centro earthquake records with a time step $\Delta{t}=0.02$ are used to simulate the seismic loads and are applied in the transverse direction, as shown in Figure \ref{Fig12}.

The cross-sectional areas $A$ of twenty-nine frames and material properties, i.e., modulus of elasticity $E$ and yield strength $f_y$, of three types of steel are considered random variables, which may have dominant effects on the failure of the tower \cite{albermani2009failure}\cite{kim2023estimation}. Table \ref{Tab_ex3} summarizes the distribution models and parameters of the thirty-five random variables. The model function $\mathcal{M}$ is the response in the transverse direction at the top of the tower, and the failure is defined as the top displacement exceeding a threshold of 0.193 \textit{m}. The finite difference method is used to calculate the gradients employing additional simulations for each variable dimension, for a physics-based model one may introduce other alternatives. (see Section \ref{Issues1} for a discussion of gradient calculation). An initial training set of 50 samples is used in the AaS-hGP method. Again, Table \ref{Tab_ex3_2} confirms that AaS-hGP successfully handles this high-dimensional reliability problem involving finite element simulations.

\begin{table}[H]
  \caption{\textbf{Distribution models and parameters of the random variables for the transmission tower}.}
  \label{Tab_ex3}
  \centering
  \begin{tabular}{c c c c}
    \toprule
    Random variables & Distribution & Mean & c.o.v. \\
    \midrule
    $A_i \, (mm^2)$ & Gaussian & $\mu_{A_i}$ & 0.07 \\
    $E_{s240} \, (Mpa)$ & Lognormal & 200,000 & 0.03 \\
    $E_{s250} \, (Mpa)$ & Lognormal & 200,000 & 0.03 \\
    $E_{s335} \, (Mpa)$ & Lognormal & 200,000 & 0.03 \\
    $f_{y,s240} \, (Mpa)$ & Lognormal & 240 & 0.05 \\
    $f_{y,s250} \, (Mpa)$ & Lognormal & 250 & 0.05 \\
    $f_{y,s335} \, (Mpa)$ & Lognormal & 335 & 0.05 \\
    \bottomrule
  \end{tabular} \\
    \raggedright{\hspace{2cm}* For $i=1,...,29$} \\
    \raggedright{\hspace{2cm}* $\mu_{A_i}$ are deterministic parameters that vary within the range of [344, 26195] \\
    \hspace{2cm} for each frame of the transmission tower structure} \\
\end{table}

\begin{table}[H]
  \caption{\textbf{Performance of AaS-hGP compared with other reliability analysis methods for the
transmission tower example}.}
  \label{Tab_ex3_2}
  \centering
  \begin{tabular}{c c c c c c}
    \toprule
    Methods & $\hat{P}_{f}$ & $N_s$ & $N_g$ & $\hat{\beta}_{g}$ & $\varepsilon_p(\%)$\\
    \midrule
    MCS & $4.50\times10^{-3}$ & $10^4$ & - & 2.61 & - \\
    AK-MCS & - & - & - & - & - \\
    \multirow{2}{8em}{\centering{ AaS-hGP\\(w/ global DoEs)}} & \multirow{2}{6em}{\centering{$8.78\times10^{-4}$}} & \multirow{2}{3em}{\centering{100}} & \multirow{2}{3em}{\centering{100}} & \multirow{2}{3em}{\centering{3.13}} & \multirow{2}{3em}{\centering{80.49}} \\
    &  &  &  &  &  \\ 
    AaS-hGP & $4.53\times10^{-3}$ & 86 & 86 & 2.61 & 0.66 \\
    \bottomrule
  \end{tabular}
\end{table}

\section{Practical issues, limitations, and future directions} \label{Issues}
\subsection{Calculation of gradients} \label{Issues1}
\noindent One practical issue that deserves attention is that the active subspace method requires gradient evaluations of a computational model. In computer-aided engineering, the gradients of physics-based models are often available in finite element software, e.g., OpenSees \cite{altoontash2004simulation}. When direct differentiation is unavailable, the numerical differentiation schemes, e.g., finite difference method, complex-valued approach, and adjoint-based strategies, can be employed to approximate the gradient using perturbations for each variable and/or equilibrium of physical equations \cite{choi2004structural}.

Recently, forward and backward automatic differentiation, also termed algorithmic differentiation, has gained wide popularity and has been considered an alternative to finite difference and direct differential methods. The automatic differentiation redefines the semantics of the gradient operator to propagate derivatives per the chain rule of differential calculus \cite{paszke2017automatic}\cite{baydin2018automatic}. It computes gradients by accumulating values during computer code execution to generate numerical derivative evaluations rather than derivative expressions, using computing resources compatible with the original model evaluations. The general-purpose automatic differentiation toolbox \cite{willkommadimat}\cite{paszke2017automatic} has been developed and applied to various engineering problems such as design optimization, computational fluid dynamics, and optimal controls; this makes gradient computations of the active subspace method easily achievable.

\subsection{Low probability problems}
\noindent In this paper, the performance of the proposed method is tested with failure probabilities down to $10^{-4}$. If an extremely small failure probability (e.g. $10^{-7}$ is of interest, the variance-reduction approaches \cite{zuev2012bayesian}\cite{papaioannou2019improved}\cite{wang2019hamiltonian} would be preferable over a direct Monte Carlo simulation (Step 3 of the algorithm in Section \ref{Algorithmsection}). The synergy between variance-reduction approaches and the proposed AaS-hGP method for high-dimensional reliability estimation is worthy of future research.

\subsection{Decision-making under constraints on reliability}
\noindent While this paper demonstrated that the proposed method was effective for high-dimensional reliability assessment, the design optimization considering these reliabilities is also essential for optimal decision-making under uncertainties. Reliability-based design optimization usually involves global optimization algorithms and sensitivities of reliability metrics with respect to design variables \cite{song2021structural}\cite{jerez2022reliability}. Thus, introducing an outer-loop stochastic optimization algorithm into the proposed AaS-hGP method will be promising for solving reliability-based design optimization problems involving high-dimensional uncertainties.

\section{Conclusions}\label{Conclusion}

\noindent This paper proposed a new adaptive dimensionality reduction-based metamodeling framework for high-dimensional reliability analysis, termed adaptive active subspace-based heteroscedastic Gaussian process (AaS-hGP). The main objective of AaS-hGP is to solve reliability problems with high-dimensional uncertainties and expensive computational models using a limited computational budget. The proposed approach identifies the optimal low-dimensional features by leveraging active subspace mapping in conjunction with hGP metamodeling in the active subspace. The proposed active learning scheme further reduces the number of training points by identifying the critical training data points significantly contributing to the failure probability.

The performance and merits of the proposed method were successfully demonstrated through several numerical examples, including a high-dimensional nonlinear mathematical function and engineering applications. In each example, the proposed method required a small number of model function and gradient evaluations to achieve accurate results. The transmission tower example demonstrated that AaS-hGP could effectively handle expensive computational models defined by the finite element method.

\section*{Acknowledgement}
\noindent
The authors are grateful to the Korea Electric Power Corporation Engineering and Construction (KEPCO E\&C) for the finite element model of the transmission tower used in this paper, co-developed with Prof. Tae-Hyung Lee at KonKuk University. The work of J. Song was supported by the National Research Foundation of Korea (NRF) Grant funded by the Korean government (NRF-2021R1A2C2003553) and the Institute of Construction and Environmental Engineering at Seoul National University. These supports are gratefully acknowledged.





\appendix
\renewcommand{\theequation}{A.\arabic{equation}}
\renewcommand{\thefigure}{A.\arabic{figure}}
\renewcommand{\thetable}{A.\arabic{table}}
\setcounter{figure}{0} 
\setcounter{table}{0} 

\section{Motivation for using active subspace as dimensionality reduction}\label{App:A}
\begin{figure}[H]
  \centering
  \includegraphics[scale=0.22] {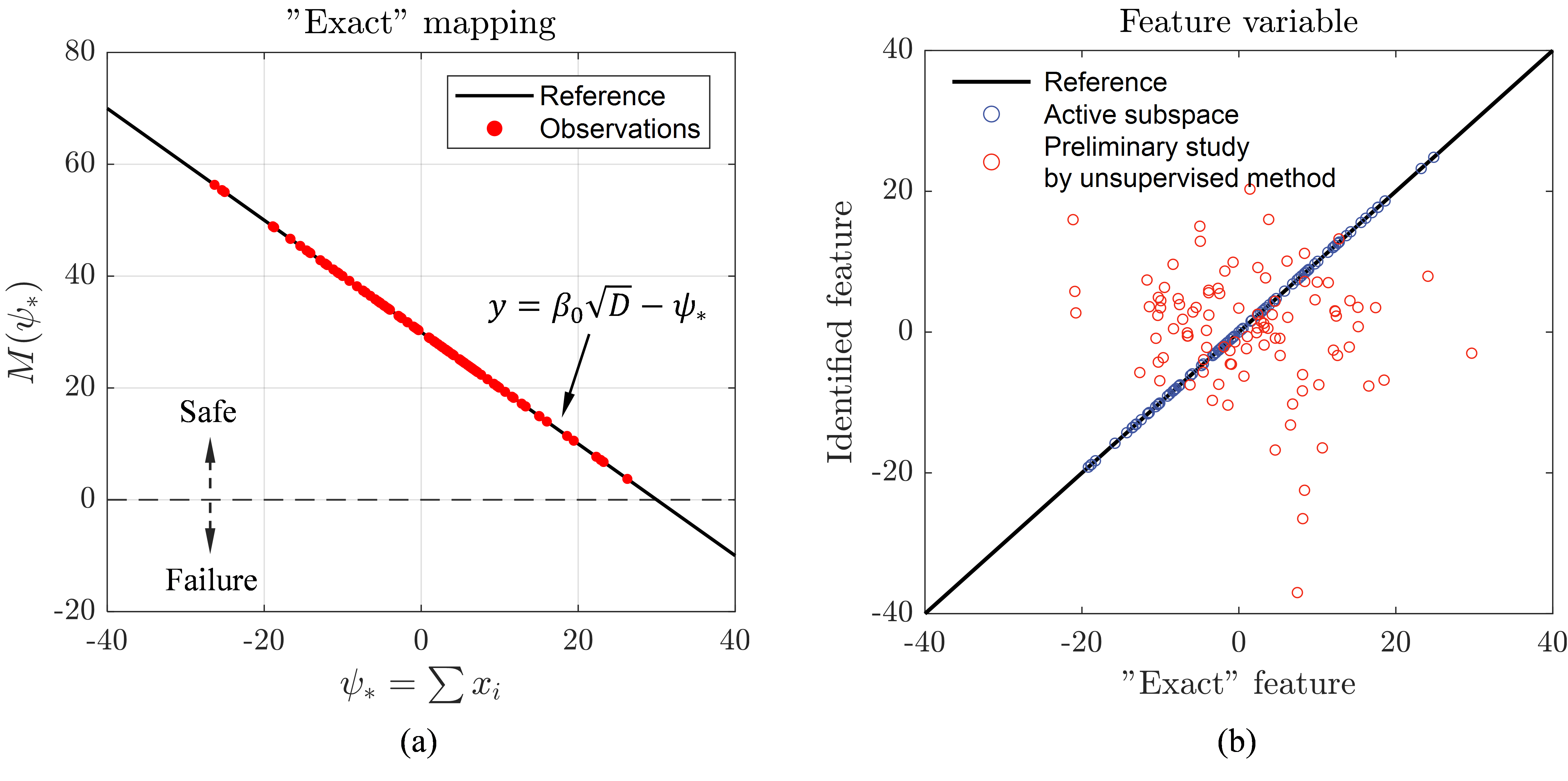}
  \caption{\textbf{(a) Illustration of the “exact” mapping and (b) scatter plot of the features obtained by the best unsupervised method and active subspace for the high-dimensional linear example}.} \label{FigA1}
\end{figure}

\noindent This appendix presents preliminary studies on various dimensionality reduction schemes. Consider the following high-dimensional linear model function:
\begin{equation}
Y = \mathcal{M}(\vect{X}) = \beta_0\sqrt{D} - \sum_{j=1}^{D}{X_j}   \, ,\label{AppendixA1}
\end{equation}
\noindent where $X_j, \, j=1,...,D$, are standard Gaussian random variables; $D$ is the dimension; and $\beta_0$ denotes the model parameter. With threshold $y_f=0$, the probability of failure is set to $P_f=\Phi({-\beta_0})$ regardless of the dimensionality of the input $\vect{X}$. Note that this model function has the “exact” one-dimensional feature mapping:
\begin{equation}
\Psi_* = \sum_{j=1}^{D}{X_j}   \, ,\label{AppendixA1_2}
\end{equation}
\noindent which can be used to predict the failure probability with perfect accuracy. The illustration of the exact feature mapping is presented in Figure \ref{FigA1}.

To numerically identify the low-dimensional feature, we test various dimensionality reduction techniques, such as PCA, kernel PCA, diffusion maps, and local linear embedding, using the dimensionality reduction toolbox developed by Van Der Maaten et al. (2009) \cite{van2009dimensionality}. Table A1 lists the investigated dimensionality reduction methods and their parameters. For each dimensionality reduction method, the optimal parameter values, i.e., $\vect{\theta}_{\mathcal{H}}$ in Eq.\eqref{DR}, are identified by minimizing the metamodeling error in Eq.\eqref{error_dr} using a global optimization algorithm. The dimensionality reduction method with the minimum metamodeling error is selected to report the results in Figure \ref{FigA1}(b).

Figure \ref{FigA1}(b) shows the scatter plot of the exact feature against the identified feature using 100 training points. The model parameter is $\beta_0=3$ and thus $P_{f,exact}=1.35\times10^{-3}$. The dimension is set to $D=100$, and the target reduced dimensionality is fixed to $d_r=1$ to check if the “exact” feature can be identified. The results show that the unsupervised techniques listed in Table \ref{Tab_A1} fail to detect the “exact” feature mapping, and the best failure probability estimate is $\hat{P}_{f}=8.20\times10^{-5}$, which is far from correct. On the other hand, the active subspace accurately solves this problem.

\begin{table}[H]
  \caption{\textbf{Dimensionality reduction methods and parameter settings investigated in the preliminary study}.}
  \label{Tab_A1}
  \centering
  \begin{tabular}{c c}
    \toprule
    Dimension reduction methods & Parameters \\
    \midrule
    PCA & None \\
    Diffusion maps & $0.5 \leq t \leq 20, \, 0.5 \leq \sigma \leq 3$ \\
    \multirow{2}{15em}{\centering{ Kernel PCA\\(with polynomial kernel)}} & \multirow{2}{12em}{\centering{$0.5 \leq a \leq 5, \, 0.5 \leq b \leq 5$}}\\
    &  \\ 
    Local linear embedding (LLE) & $3 \leq k \leq 10$ \\
    Hessian LLE & $3 \leq k \leq 10$ \\
    Maximum variance unfolding & $5 \leq k \leq 15$ \\
    Laplacian eigenmaps & $5 \leq k \leq 15, \, \sigma=1 $\\
    \bottomrule
  \end{tabular} \\
\end{table}

\renewcommand{\theequation}{B.\arabic{equation}}

\section{Basic theories of the homoscedastic GP model}\label{App:B}

\noindent The central assumption of a Gaussian process model is that the response at the input $\vect{x}$, $y(\vect{x})$ is the realization of a Gaussian process \cite{williams2006gaussian}, denoted as 
\begin{equation}
f(\vect{x}) \sim GP\left(\mu_{f}(\vect{x}),k_f\left({\vect{x},\vect{x}';\vect{\theta_f}}\right)\right)   \, ,\label{AppendixB1_1}
\end{equation}
\noindent where $\mu_{f}(\vect{x})=\E{y(\vect{x})}$ is the mean function; $k_f(\vect{x},\vect{x}')=\E{(f(\vect{x})-\mu_{f}(\vect{x}))(f(\vect{x}')-\mu_{f}(\vect{x}'))}$ is the covariance function (or “kernel” function); and $\vect{\theta_f}$ is a set of parameters that characterize the process, often termed hyperparameters.

GP usually assumes that the observations incorporate homoscedastic Gaussian noise $\varepsilon$, which can be expressed as
\begin{equation}
\mathcal{Y}=f(\vect{x})+\varepsilon   \, ,\label{AppendixB1_2}
\end{equation}
\noindent where the Gaussian noises $\varepsilon$ are statistically independent of each other and identically distributed with constant variance $\sigma_n^2$. Given $n$ training samples of inputs $\vect{x}_{\mathcal{D}}=[\vect{x}_1,...,\vect{x}_n]^T$ and the corresponding noisy observations $\vect{\mathcal{Y}}_{\mathcal{D}}=[\mathcal{Y}(\vect{x}_1),...,\mathcal{Y}(\vect{x}_n)]^T$, the optimal hyperparameter estimates, $\hat{\vect{\Theta}}$ can be obtained by the maximum likelihood estimation (MLE) method \cite{williams2006gaussian}, i.e.,
\begin{equation}
\hat{\vect{\Theta}} = \mathop{\arg\max}_{\vect{\Theta}}{\ln{p(\vect{\mathcal{Y}}_{\mathcal{D}} | \vect{x}_{\mathcal{D}},\vect{\Theta})}} \,,\,\,\,\,   \, 
\label{AppendixB1_3}
\end{equation}
\begin{equation}
\ln{p(\vect{\mathcal{Y}}_{\mathcal{D}} | \vect{x}_{\mathcal{D}},\vect{\Theta})} = -\frac{1}{2}\vect{\mathcal{Y}}_{\mathcal{D}}^T(\vect{K_f} + \sigma_n^2\vect{I})^{-1}\vect{\mathcal{Y}}_{\mathcal{D}} - \frac{1}{2} \ln{|\vect{K_f} + \sigma_n^2\vect{I}|} - \frac{n}{2} \ln{2\pi}
\,, \label{AppendixB1_4}
\end{equation}
\noindent where $\vect{K_f}$ is the covariance matrix whose element is determined as $K_{f_{i,j}}=k_f(\vect{x}_i,\vect{x}_j), \, i,j=1,...,n$; and $\vect{I}$ is the $(n \times n)$ identity matrix.

Then, the GP model can predict the responses at new input points with optimal hyperparameters. In detail, with the zero-mean function, the response prediction $\mu_{\hat{Y},GP}\left(\cdot\right)$ and the corresponding prediction variance $\sigma^2_{\hat{Y},GP}\left(\cdot\right)$ at a new point $\vect{x}_*$ can be computed as
\begin{equation}
\mu_{\hat{Y},GP}\left(\vect{x}_*\right) = \vect{k}_{f_*}^T(\vect{K_f} + \sigma_n^2\vect{I})^{-1}\vect{\mathcal{Y}}_{\mathcal{D}}     \,, \label{AppendixB1_5}
\end{equation}
\begin{equation}
\sigma^2_{\hat{Y},GP}\left(\vect{x}_*\right) = k_{f_{**}} - \vect{k}_{f_*}^T(\vect{K_f} + \sigma_n^2\vect{I})^{-1}\vect{k}_{f_*}     \, ,\label{AppendixB1_6}
\end{equation}
\noindent where $\vect{k}_{f_*} = [k_f(\vect{x}_*,\vect{x}_1),...,k_f(\vect{x}_*,\vect{x}_n)]^T$ includes the covariances between the prediction location $\vect{x}_*$ and $n$ observed points $\vect{x}_{\mathcal{D}}$; and $k_{f_{**}} = k_f(\vect{x}_*,\vect{x}_*)$. Thus, the GP model not only provides the mean estimate $\mu_{\hat{Y},GP}\left(\vect{x}_*\right)$ but also quantifies the uncertainty of the prediction by $\sigma^2_{\hat{Y},GP}\left(\vect{x}_*\right)$ under the homoscedastic noise assumptions.

\bibliography{AaShGP}

\begin{thebibliography}{10}

\bibitem{au2003subset}
Siu-Kui Au and James~L Beck.
\newblock Subset simulation and its application to seismic risk based on
  dynamic analysis.
\newblock {\em Journal of Engineering Mechanics}, 129(8):901--917, 2003.

\bibitem{zuev2012bayesian}
Konstantin~M Zuev, James~L Beck, Siu-Kui Au, and Lambros~S Katafygiotis.
\newblock Bayesian post-processor and other enhancements of subset simulation
  for estimating failure probabilities in high dimensions.
\newblock {\em Computers \& structures}, 92:283--296, 2012.

\bibitem{wang2019hamiltonian}
Ziqi Wang, Marco Broccardo, and Junho Song.
\newblock Hamiltonian monte carlo methods for subset simulation in reliability
  analysis.
\newblock {\em Structural Safety}, 76:51--67, 2019.

\bibitem{wang2016cross}
Ziqi Wang and Junho Song.
\newblock Cross-entropy-based adaptive importance sampling using von
  mises-fisher mixture for high dimensional reliability analysis.
\newblock {\em Structural Safety}, 59:42--52, 2016.

\bibitem{papaioannou2019improved}
Iason Papaioannou, Sebastian Geyer, and Daniel Straub.
\newblock Improved cross entropy-based importance sampling with a flexible
  mixture model.
\newblock {\em Reliability Engineering \& System Safety}, 191:106564, 2019.

\bibitem{lataniotis2020extending}
Christos Lataniotis, Stefano Marelli, and Bruno Sudret.
\newblock Extending classical surrogate modeling to high dimensions through
  supervised dimensionality reduction: a data-driven approach.
\newblock {\em International Journal for Uncertainty Quantification}, 10(1),
  2020.

\bibitem{kim2021quantile}
Jungho Kim and Junho Song.
\newblock Quantile surrogates and sensitivity by adaptive gaussian process for
  efficient reliability-based design optimization.
\newblock {\em Mechanical Systems and Signal Processing}, 161:107962, 2021.

\bibitem{kontolati2022survey}
Katiana Kontolati, Dimitrios Loukrezis, Dimitrios~G Giovanis, Lohit Vandanapu,
  and Michael~D Shields.
\newblock A survey of unsupervised learning methods for high-dimensional
  uncertainty quantification in black-box-type problems.
\newblock {\em Journal of Computational Physics}, 464:111313, 2022.

\bibitem{kalogeris2020diffusion}
I~Kalogeris and V~Papadopoulos.
\newblock Diffusion maps-based surrogate modeling: An alternative machine
  learning approach.
\newblock {\em International Journal for Numerical Methods in Engineering},
  121(4):602--620, 2020.

\bibitem{giovanis2018uncertainty}
Dimitris~G Giovanis and Michael~D Shields.
\newblock Uncertainty quantification for complex systems with very high
  dimensional response using grassmann manifold variations.
\newblock {\em Journal of Computational Physics}, 364:393--415, 2018.

\bibitem{giovanis2020data}
Dimitris~G Giovanis and Michael~D Shields.
\newblock Data-driven surrogates for high dimensional models using gaussian
  process regression on the grassmann manifold.
\newblock {\em Computer Methods in Applied Mechanics and Engineering},
  370:113269, 2020.

\bibitem{dos2022grassmannian}
Ketson~R Dos~Santos, Dimitrios~G Giovanis, and Michael~D Shields.
\newblock Grassmannian diffusion maps--based dimension reduction and
  classification for high-dimensional data.
\newblock {\em SIAM Journal on Scientific Computing}, 44(2):B250--B274, 2022.

\bibitem{li2020deep}
Mingyang Li and Zequn Wang.
\newblock Deep learning for high-dimensional reliability analysis.
\newblock {\em Mechanical Systems and Signal Processing}, 139:106399, 2020.

\bibitem{jiang2021recursive}
Zhong-ming Jiang, De-Cheng Feng, Hao Zhou, and Wei-Feng Tao.
\newblock A recursive dimension-reduction method for high-dimensional
  reliability analysis with rare failure event.
\newblock {\em Reliability Engineering \& System Safety}, 213:107710, 2021.

\bibitem{constantine2014active}
Paul~G Constantine, Eric Dow, and Qiqi Wang.
\newblock Active subspace methods in theory and practice: applications to
  kriging surfaces.
\newblock {\em SIAM Journal on Scientific Computing}, 36(4):A1500--A1524, 2014.

\bibitem{jiang2017high}
Zhongming Jiang and Jie Li.
\newblock High dimensional structural reliability with dimension reduction.
\newblock {\em Structural Safety}, 69:35--46, 2017.

\bibitem{zhou2021active}
Tong Zhou and Yongbo Peng.
\newblock Active learning and active subspace enhancement for pdem-based
  high-dimensional reliability analysis.
\newblock {\em Structural Safety}, 88:102026, 2021.

\bibitem{navaneeth2022surrogate}
N~Navaneeth and Souvik Chakraborty.
\newblock Surrogate assisted active subspace and active subspace assisted
  surrogate—a new paradigm for high dimensional structural reliability
  analysis.
\newblock {\em Computer Methods in Applied Mechanics and Engineering},
  389:114374, 2022.

\bibitem{der2022structural}
Armen Der~Kiureghian.
\newblock {\em Structural and system reliability}.
\newblock Cambridge University Press, 2022.

\bibitem{song2021structural}
Junho Song, Won-Hee Kang, Young-Joo Lee, and Junho Chun.
\newblock Structural system reliability: Overview of theories and applications
  to optimization.
\newblock {\em ASCE-ASME Journal of Risk and Uncertainty in Engineering
  Systems, Part A: Civil Engineering}, 7(2):03121001, 2021.

\bibitem{alibrandi2014response}
Umberto Alibrandi.
\newblock A response surface method for stochastic dynamic analysis.
\newblock {\em Reliability Engineering \& System Safety}, 126:44--53, 2014.

\bibitem{kim2023estimation}
Jungho Kim, Sang-ri Yi, and Junho Song.
\newblock Estimation of first-passage probability under stochastic wind
  excitations by active-learning-based heteroscedastic gaussian process.
\newblock {\em Structural Safety}, 100:102268, 2023.

\bibitem{van2009dimensionality}
Laurens Van Der~Maaten, Eric Postma, Jaap Van~den Herik, et~al.
\newblock Dimensionality reduction: a comparative.
\newblock {\em J Mach Learn Res}, 10(66-71):13, 2009.

\bibitem{blatman2011adaptive}
G{\'e}raud Blatman and Bruno Sudret.
\newblock Adaptive sparse polynomial chaos expansion based on least angle
  regression.
\newblock {\em Journal of computational Physics}, 230(6):2345--2367, 2011.

\bibitem{zhang2019accelerating}
Jize Zhang and Alexandros~A Taflanidis.
\newblock Accelerating mcmc via kriging-based adaptive independent proposals
  and delayed rejection.
\newblock {\em Computer Methods in Applied Mechanics and Engineering},
  355:1124--1147, 2019.

\bibitem{kim2020probability}
Jungho Kim and Junho Song.
\newblock Probability-adaptive kriging in n-ball (pak-bn) for reliability
  analysis.
\newblock {\em Structural Safety}, 85:101924, 2020.

\bibitem{nguyen2019ten}
Lan~Huong Nguyen and Susan Holmes.
\newblock Ten quick tips for effective dimensionality reduction.
\newblock {\em PLoS computational biology}, 15(6):e1006907, 2019.

\bibitem{lazaro2013retrieval}
Miguel L{\'a}zaro-Gredilla, Michalis~K Titsias, Jochem Verrelst, and Gustavo
  Camps-Valls.
\newblock Retrieval of biophysical parameters with heteroscedastic gaussian
  processes.
\newblock {\em IEEE Geoscience and Remote Sensing Letters}, 11(4):838--842,
  2013.

\bibitem{kim2021clustering}
Taeyong Kim, Oh-Sung Kwon, and Junho Song.
\newblock Clustering-based adaptive ground motion selection algorithm for
  efficient estimation of structural fragilities.
\newblock {\em Earthquake Engineering \& Structural Dynamics},
  50(6):1755--1776, 2021.

\bibitem{marrel2008efficient}
Amandine Marrel, Bertrand Iooss, Fran{\c{c}}ois Van~Dorpe, and Elena Volkova.
\newblock An efficient methodology for modeling complex computer codes with
  gaussian processes.
\newblock {\em Computational Statistics \& Data Analysis}, 52(10):4731--4744,
  2008.

\bibitem{echard2011ak}
Benjamin Echard, Nicolas Gayton, and Maurice Lemaire.
\newblock Ak-mcs: an active learning reliability method combining kriging and
  monte carlo simulation.
\newblock {\em Structural Safety}, 33(2):145--154, 2011.

\bibitem{rahami2008sizing}
H~Rahami, A~Kaveh, and Y~Gholipour.
\newblock Sizing, geometry and topology optimization of trusses via force
  method and genetic algorithm.
\newblock {\em Engineering Structures}, 30(9):2360--2369, 2008.

\bibitem{park2016seismic}
Hyo-Sang Park, Byung~Ho Choi, Jung~Joong Kim, and Tae-Hyung Lee.
\newblock Seismic performance evaluation of high voltage transmission towers in
  south korea.
\newblock {\em KSCE Journal of Civil Engineering}, 20:2499--2505, 2016.

\bibitem{albermani2009failure}
Faris Albermani, Sritawat Kitipornchai, and Ricky~WK Chan.
\newblock Failure analysis of transmission towers.
\newblock {\em Engineering failure analysis}, 16(6):1922--1928, 2009.

\bibitem{altoontash2004simulation}
Arash Altoontash.
\newblock {\em Simulation and damage models for performance assessment of
  reinforced concrete beam-column joints}.
\newblock Stanford university, 2004.

\bibitem{choi2004structural}
Kyung~K Choi and Nam-Ho Kim.
\newblock {\em Structural sensitivity analysis and optimization 1: linear
  systems}.
\newblock Springer Science \& Business Media, 2004.

\bibitem{paszke2017automatic}
Adam Paszke, Sam Gross, Soumith Chintala, Gregory Chanan, Edward Yang, Zachary
  DeVito, Zeming Lin, Alban Desmaison, Luca Antiga, and Adam Lerer.
\newblock Automatic differentiation in pytorch.
\newblock 2017.

\bibitem{baydin2018automatic}
Atilim~Gunes Baydin, Barak~A Pearlmutter, Alexey~Andreyevich Radul, and
  Jeffrey~Mark Siskind.
\newblock Automatic differentiation in machine learning: a survey.
\newblock {\em Journal of Marchine Learning Research}, 18:1--43, 2018.

\bibitem{willkommadimat}
J~Willkomm and A~Vehreschild.
\newblock The adimat handbook, 2013.
\newblock {\em URL http://adimat. sc. informatik. tu-darmstadt. de/doc}.

\bibitem{jerez2022reliability}
DJ~Jerez, HA~Jensen, and M~Beer.
\newblock Reliability-based design optimization of structural systems under
  stochastic excitation: An overview.
\newblock {\em Mechanical Systems and Signal Processing}, 166:108397, 2022.

\bibitem{williams2006gaussian}
Christopher~KI Williams and Carl~Edward Rasmussen.
\newblock {\em Gaussian processes for machine learning}, volume~2.
\newblock MIT press Cambridge, MA, 2006.

\end{thebibliography}







\end{document}